\pdfoutput=1
\documentclass[11pt,twoside,a4paper,pdftex,cmspaper,final,collab]{cms-tdr}

\begin{document}\cmsNoteHeader{CFT-09-001}
%
%
%

%
%
\hyphenation{env-iron-men-tal}
\hyphenation{had-ron-i-za-tion}
\hyphenation{cal-or-i-me-ter}
\hyphenation{de-vices}
%
%
\RCS$Revision: 1.15 $
\RCS$Date: 2010/02/02 12:28:06 $
\RCS$Name:  $
%
%
%

\newcommand {\etal}{\mbox{et al.}\xspace} 
\newcommand {\ie}{\mbox{i.e.}\xspace}     
\newcommand {\eg}{\mbox{e.g.}\xspace}     
\newcommand {\etc}{\mbox{etc.}\xspace}     
\newcommand {\vs}{\mbox{\sl vs.}\xspace}      
\newcommand {\mdash}{\ensuremath{\mathrm{-}}} 

\newcommand {\Lone}{Level-1\xspace} 
\newcommand {\Ltwo}{Level-2\xspace}
\newcommand {\Lthree}{Level-3\xspace}

\providecommand{\ACERMC} {\textsc{AcerMC}\xspace}
\providecommand{\ALPGEN} {{\textsc{alpgen}}\xspace}
\providecommand{\CHARYBDIS} {{\textsc{charybdis}}\xspace}
\providecommand{\CMKIN} {\textsc{cmkin}\xspace}
\providecommand{\CMSIM} {{\textsc{cmsim}}\xspace}
\providecommand{\CMSSW} {{\textsc{cmssw}}\xspace}
\providecommand{\COBRA} {{\textsc{cobra}}\xspace}
\providecommand{\COCOA} {{\textsc{cocoa}}\xspace}
\providecommand{\COMPHEP} {\textsc{CompHEP}\xspace}
\providecommand{\EVTGEN} {{\textsc{evtgen}}\xspace}
\providecommand{\FAMOS} {{\textsc{famos}}\xspace}
\providecommand{\GARCON} {\textsc{garcon}\xspace}
\providecommand{\GARFIELD} {{\textsc{garfield}}\xspace}
\providecommand{\GEANE} {{\textsc{geane}}\xspace}
\providecommand{\GEANTfour} {{\textsc{geant4}}\xspace}
\providecommand{\GEANTthree} {{\textsc{geant3}}\xspace}
\providecommand{\GEANT} {{\textsc{geant}}\xspace}
\providecommand{\HDECAY} {\textsc{hdecay}\xspace}
\providecommand{\HERWIG} {{\textsc{herwig}}\xspace}
\providecommand{\HIGLU} {{\textsc{higlu}}\xspace}
\providecommand{\HIJING} {{\textsc{hijing}}\xspace}
\providecommand{\IGUANA} {\textsc{iguana}\xspace}
\providecommand{\ISAJET} {{\textsc{isajet}}\xspace}
\providecommand{\ISAPYTHIA} {{\textsc{isapythia}}\xspace}
\providecommand{\ISASUGRA} {{\textsc{isasugra}}\xspace}
\providecommand{\ISASUSY} {{\textsc{isasusy}}\xspace}
\providecommand{\ISAWIG} {{\textsc{isawig}}\xspace}
\providecommand{\MADGRAPH} {\textsc{MadGraph}\xspace}
\providecommand{\MCATNLO} {\textsc{mc@nlo}\xspace}
\providecommand{\MCFM} {\textsc{mcfm}\xspace}
\providecommand{\MILLEPEDE} {{\textsc{millepede}}\xspace}
\providecommand{\ORCA} {{\textsc{orca}}\xspace}
\providecommand{\OSCAR} {{\textsc{oscar}}\xspace}
\providecommand{\PHOTOS} {\textsc{photos}\xspace}
\providecommand{\PROSPINO} {\textsc{prospino}\xspace}
\providecommand{\PYTHIA} {{\textsc{pythia}}\xspace}
\providecommand{\SHERPA} {{\textsc{sherpa}}\xspace}
\providecommand{\TAUOLA} {\textsc{tauola}\xspace}
\providecommand{\TOPREX} {\textsc{TopReX}\xspace}
\providecommand{\XDAQ} {{\textsc{xdaq}}\xspace}

\newcommand {\DZERO}{D\O\xspace}     


\newcommand{\de}{\ensuremath{^\circ}}
\newcommand{\ten}[1]{\ensuremath{\times \text{10}^\text{#1}}}
\newcommand{\unit}[1]{\ensuremath{\text{\,#1}}\xspace}
\newcommand{\mum}{\ensuremath{\,\mu\text{m}}\xspace}
\newcommand{\micron}{\ensuremath{\,\mu\text{m}}\xspace}
\newcommand{\cm}{\ensuremath{\,\text{cm}}\xspace}
\newcommand{\mm}{\ensuremath{\,\text{mm}}\xspace}
\newcommand{\mus}{\ensuremath{\,\mu\text{s}}\xspace}
\newcommand{\keV}{\ensuremath{\,\text{ke\hspace{-.08em}V}}\xspace}
\newcommand{\MeV}{\ensuremath{\,\text{Me\hspace{-.08em}V}}\xspace}
\newcommand{\GeV}{\ensuremath{\,\text{Ge\hspace{-.08em}V}}\xspace}
\newcommand{\TeV}{\ensuremath{\,\text{Te\hspace{-.08em}V}}\xspace}
\newcommand{\PeV}{\ensuremath{\,\text{Pe\hspace{-.08em}V}}\xspace}
\newcommand{\keVc}{\ensuremath{{\,\text{ke\hspace{-.08em}V\hspace{-0.16em}/\hspace{-0.08em}c}}}\xspace}
\newcommand{\MeVc}{\ensuremath{{\,\text{Me\hspace{-.08em}V\hspace{-0.16em}/\hspace{-0.08em}c}}}\xspace}
\newcommand{\GeVc}{\ensuremath{{\,\text{Ge\hspace{-.08em}V\hspace{-0.16em}/\hspace{-0.08em}c}}}\xspace}
\newcommand{\TeVc}{\ensuremath{{\,\text{Te\hspace{-.08em}V\hspace{-0.16em}/\hspace{-0.08em}c}}}\xspace}
\newcommand{\keVcc}{\ensuremath{{\,\text{ke\hspace{-.08em}V\hspace{-0.16em}/\hspace{-0.08em}c}^\text{2}}}\xspace}
\newcommand{\MeVcc}{\ensuremath{{\,\text{Me\hspace{-.08em}V\hspace{-0.16em}/\hspace{-0.08em}c}^\text{2}}}\xspace}
\newcommand{\GeVcc}{\ensuremath{{\,\text{Ge\hspace{-.08em}V\hspace{-0.16em}/\hspace{-0.08em}c}^\text{2}}}\xspace}
\newcommand{\TeVcc}{\ensuremath{{\,\text{Te\hspace{-.08em}V\hspace{-0.16em}/\hspace{-0.08em}c}^\text{2}}}\xspace}

\newcommand{\pbinv} {\mbox{\ensuremath{\,\text{pb}^\text{$-$1}}}\xspace}
\newcommand{\fbinv} {\mbox{\ensuremath{\,\text{fb}^\text{$-$1}}}\xspace}
\newcommand{\nbinv} {\mbox{\ensuremath{\,\text{nb}^\text{$-$1}}}\xspace}
\newcommand{\percms}{\ensuremath{\,\text{cm}^\text{$-$2}\,\text{s}^\text{$-$1}}\xspace}
\newcommand{\lumi}{\ensuremath{\mathcal{L}}\xspace}
\newcommand{\Lumi}{\ensuremath{\mathcal{L}}\xspace}
%
\newcommand{\LvLow}  {\ensuremath{\mathcal{L}=\text{10}^\text{32}\,\text{cm}^\text{$-$2}\,\text{s}^\text{$-$1}}\xspace}
\newcommand{\LLow}   {\ensuremath{\mathcal{L}=\text{10}^\text{33}\,\text{cm}^\text{$-$2}\,\text{s}^\text{$-$1}}\xspace}
\newcommand{\lowlumi}{\ensuremath{\mathcal{L}=\text{2}\times \text{10}^\text{33}\,\text{cm}^\text{$-$2}\,\text{s}^\text{$-$1}}\xspace}
\newcommand{\LMed}   {\ensuremath{\mathcal{L}=\text{2}\times \text{10}^\text{33}\,\text{cm}^\text{$-$2}\,\text{s}^\text{$-$1}}\xspace}
\newcommand{\LHigh}  {\ensuremath{\mathcal{L}=\text{10}^\text{34}\,\text{cm}^\text{$-$2}\,\text{s}^\text{$-$1}}\xspace}
\newcommand{\hilumi} {\ensuremath{\mathcal{L}=\text{10}^\text{34}\,\text{cm}^\text{$-$2}\,\text{s}^\text{$-$1}}\xspace}


\newcommand{\zp}{\ensuremath{\mathrm{Z}^\prime}\xspace}


\newcommand{\kt}{\ensuremath{k_{\mathrm{T}}}\xspace}
\newcommand{\BC}{\ensuremath{{B_{\mathrm{c}}}}\xspace}
\newcommand{\bbarc}{\ensuremath{{\overline{b}c}}\xspace}
\newcommand{\bbbar}{\ensuremath{{b\overline{b}}}\xspace}
\newcommand{\ccbar}{\ensuremath{{c\overline{c}}}\xspace}
\newcommand{\JPsi}{\ensuremath{{J}/\psi}\xspace}
\newcommand{\bspsiphi}{\ensuremath{B_s \to \JPsi\, \phi}\xspace}
\newcommand{\AFB}{\ensuremath{A_\mathrm{FB}}\xspace}
\newcommand{\EE}{\ensuremath{e^+e^-}\xspace}
\newcommand{\MM}{\ensuremath{\mu^+\mu^-}\xspace}
\newcommand{\TT}{\ensuremath{\tau^+\tau^-}\xspace}
\newcommand{\wangle}{\ensuremath{\sin^{2}\theta_{\mathrm{eff}}^\mathrm{lept}(M^2_\mathrm{Z})}\xspace}
\newcommand{\ttbar}{\ensuremath{{t\overline{t}}}\xspace}
\newcommand{\stat}{\ensuremath{\,\text{(stat.)}}\xspace}
\newcommand{\syst}{\ensuremath{\,\text{(syst.)}}\xspace}

\newcommand{\HGG}{\ensuremath{\mathrm{H}\to\gamma\gamma}}
\newcommand{\gev}{\GeV}
\newcommand{\GAMJET}{\ensuremath{\gamma + \mathrm{jet}}}
\newcommand{\PPTOJETS}{\ensuremath{\mathrm{pp}\to\mathrm{jets}}}
\newcommand{\PPTOGG}{\ensuremath{\mathrm{pp}\to\gamma\gamma}}
\newcommand{\PPTOGAMJET}{\ensuremath{\mathrm{pp}\to\gamma +
\mathrm{jet}
}}
\newcommand{\MH}{\ensuremath{\mathrm{M_{\mathrm{H}}}}}
\newcommand{\RNINE}{\ensuremath{\mathrm{R}_\mathrm{9}}}
\newcommand{\DR}{\ensuremath{\Delta\mathrm{R}}}


\newcommand{\PT}{\ensuremath{p_{\mathrm{T}}}\xspace}
\newcommand{\pt}{\ensuremath{p_{\mathrm{T}}}\xspace}
\newcommand{\ET}{\ensuremath{E_{\mathrm{T}}}\xspace}
\newcommand{\HT}{\ensuremath{H_{\mathrm{T}}}\xspace}
\newcommand{\et}{\ensuremath{E_{\mathrm{T}}}\xspace}
\newcommand{\Em}{\ensuremath{E\!\!\!/}\xspace}
\newcommand{\Pm}{\ensuremath{p\!\!\!/}\xspace}
\newcommand{\PTm}{\ensuremath{{p\!\!\!/}_{\mathrm{T}}}\xspace}
\newcommand{\ETm}{\ensuremath{E_{\mathrm{T}}^{\mathrm{miss}}}\xspace}
\newcommand{\MET}{\ensuremath{E_{\mathrm{T}}^{\mathrm{miss}}}\xspace}
\newcommand{\ETmiss}{\ensuremath{E_{\mathrm{T}}^{\mathrm{miss}}}\xspace}
\newcommand{\VEtmiss}{\ensuremath{{\vec E}_{\mathrm{T}}^{\mathrm{miss}}}\xspace}

%

\newcommand{\ga}{\ensuremath{\gtrsim}}
\newcommand{\la}{\ensuremath{\lesssim}}
\newcommand{\swsq}{\ensuremath{\sin^2\theta_W}\xspace}
\newcommand{\cwsq}{\ensuremath{\cos^2\theta_W}\xspace}
\newcommand{\tanb}{\ensuremath{\tan\beta}\xspace}
\newcommand{\tanbsq}{\ensuremath{\tan^{2}\beta}\xspace}
\newcommand{\sidb}{\ensuremath{\sin 2\beta}\xspace}
\newcommand{\alpS}{\ensuremath{\alpha_S}\xspace}
\newcommand{\alpt}{\ensuremath{\tilde{\alpha}}\xspace}

\newcommand{\QL}{\ensuremath{Q_L}\xspace}
\newcommand{\sQ}{\ensuremath{\tilde{Q}}\xspace}
\newcommand{\sQL}{\ensuremath{\tilde{Q}_L}\xspace}
\newcommand{\ULC}{\ensuremath{U_L^C}\xspace}
\newcommand{\sUC}{\ensuremath{\tilde{U}^C}\xspace}
\newcommand{\sULC}{\ensuremath{\tilde{U}_L^C}\xspace}
\newcommand{\DLC}{\ensuremath{D_L^C}\xspace}
\newcommand{\sDC}{\ensuremath{\tilde{D}^C}\xspace}
\newcommand{\sDLC}{\ensuremath{\tilde{D}_L^C}\xspace}
\newcommand{\LL}{\ensuremath{L_L}\xspace}
\newcommand{\sL}{\ensuremath{\tilde{L}}\xspace}
\newcommand{\sLL}{\ensuremath{\tilde{L}_L}\xspace}
\newcommand{\ELC}{\ensuremath{E_L^C}\xspace}
\newcommand{\sEC}{\ensuremath{\tilde{E}^C}\xspace}
\newcommand{\sELC}{\ensuremath{\tilde{E}_L^C}\xspace}
\newcommand{\sEL}{\ensuremath{\tilde{E}_L}\xspace}
\newcommand{\sER}{\ensuremath{\tilde{E}_R}\xspace}
\newcommand{\sFer}{\ensuremath{\tilde{f}}\xspace}
\newcommand{\sQua}{\ensuremath{\tilde{q}}\xspace}
\newcommand{\sUp}{\ensuremath{\tilde{u}}\xspace}
\newcommand{\suL}{\ensuremath{\tilde{u}_L}\xspace}
\newcommand{\suR}{\ensuremath{\tilde{u}_R}\xspace}
\newcommand{\sDw}{\ensuremath{\tilde{d}}\xspace}
\newcommand{\sdL}{\ensuremath{\tilde{d}_L}\xspace}
\newcommand{\sdR}{\ensuremath{\tilde{d}_R}\xspace}
\newcommand{\sTop}{\ensuremath{\tilde{t}}\xspace}
\newcommand{\stL}{\ensuremath{\tilde{t}_L}\xspace}
\newcommand{\stR}{\ensuremath{\tilde{t}_R}\xspace}
\newcommand{\stone}{\ensuremath{\tilde{t}_1}\xspace}
\newcommand{\sttwo}{\ensuremath{\tilde{t}_2}\xspace}
\newcommand{\sBot}{\ensuremath{\tilde{b}}\xspace}
\newcommand{\sbL}{\ensuremath{\tilde{b}_L}\xspace}
\newcommand{\sbR}{\ensuremath{\tilde{b}_R}\xspace}
\newcommand{\sbone}{\ensuremath{\tilde{b}_1}\xspace}
\newcommand{\sbtwo}{\ensuremath{\tilde{b}_2}\xspace}
\newcommand{\sLep}{\ensuremath{\tilde{l}}\xspace}
\newcommand{\sLepC}{\ensuremath{\tilde{l}^C}\xspace}
\newcommand{\sEl}{\ensuremath{\tilde{e}}\xspace}
\newcommand{\sElC}{\ensuremath{\tilde{e}^C}\xspace}
\newcommand{\seL}{\ensuremath{\tilde{e}_L}\xspace}
\newcommand{\seR}{\ensuremath{\tilde{e}_R}\xspace}
\newcommand{\snL}{\ensuremath{\tilde{\nu}_L}\xspace}
\newcommand{\sMu}{\ensuremath{\tilde{\mu}}\xspace}
\newcommand{\sNu}{\ensuremath{\tilde{\nu}}\xspace}
\newcommand{\sTau}{\ensuremath{\tilde{\tau}}\xspace}
\newcommand{\Glu}{\ensuremath{g}\xspace}
\newcommand{\sGlu}{\ensuremath{\tilde{g}}\xspace}
\newcommand{\Wpm}{\ensuremath{W^{\pm}}\xspace}
\newcommand{\sWpm}{\ensuremath{\tilde{W}^{\pm}}\xspace}
\newcommand{\Wz}{\ensuremath{W^{0}}\xspace}
\newcommand{\sWz}{\ensuremath{\tilde{W}^{0}}\xspace}
\newcommand{\sWino}{\ensuremath{\tilde{W}}\xspace}
\newcommand{\Bz}{\ensuremath{B^{0}}\xspace}
\newcommand{\sBz}{\ensuremath{\tilde{B}^{0}}\xspace}
\newcommand{\sBino}{\ensuremath{\tilde{B}}\xspace}
\newcommand{\Zz}{\ensuremath{Z^{0}}\xspace}
\newcommand{\sZino}{\ensuremath{\tilde{Z}^{0}}\xspace}
\newcommand{\sGam}{\ensuremath{\tilde{\gamma}}\xspace}
\newcommand{\chiz}{\ensuremath{\tilde{\chi}^{0}}\xspace}
\newcommand{\chip}{\ensuremath{\tilde{\chi}^{+}}\xspace}
\newcommand{\chim}{\ensuremath{\tilde{\chi}^{-}}\xspace}
\newcommand{\chipm}{\ensuremath{\tilde{\chi}^{\pm}}\xspace}
\newcommand{\Hone}{\ensuremath{H_{d}}\xspace}
\newcommand{\sHone}{\ensuremath{\tilde{H}_{d}}\xspace}
\newcommand{\Htwo}{\ensuremath{H_{u}}\xspace}
\newcommand{\sHtwo}{\ensuremath{\tilde{H}_{u}}\xspace}
\newcommand{\sHig}{\ensuremath{\tilde{H}}\xspace}
\newcommand{\sHa}{\ensuremath{\tilde{H}_{a}}\xspace}
\newcommand{\sHb}{\ensuremath{\tilde{H}_{b}}\xspace}
\newcommand{\sHpm}{\ensuremath{\tilde{H}^{\pm}}\xspace}
\newcommand{\hz}{\ensuremath{h^{0}}\xspace}
\newcommand{\Hz}{\ensuremath{H^{0}}\xspace}
\newcommand{\Az}{\ensuremath{A^{0}}\xspace}
\newcommand{\Hpm}{\ensuremath{H^{\pm}}\xspace}
\newcommand{\sGra}{\ensuremath{\tilde{G}}\xspace}
\newcommand{\mtil}{\ensuremath{\tilde{m}}\xspace}
\newcommand{\rpv}{\ensuremath{\rlap{\kern.2em/}R}\xspace}
\newcommand{\LLE}{\ensuremath{LL\bar{E}}\xspace}
\newcommand{\LQD}{\ensuremath{LQ\bar{D}}\xspace}
\newcommand{\UDD}{\ensuremath{\overline{UDD}}\xspace}
\newcommand{\Lam}{\ensuremath{\lambda}\xspace}
\newcommand{\Lamp}{\ensuremath{\lambda'}\xspace}
\newcommand{\Lampp}{\ensuremath{\lambda''}\xspace}
\newcommand{\spinbd}[2]{\ensuremath{\bar{#1}_{\dot{#2}}}\xspace}

\newcommand{\MD}{\ensuremath{{M_\mathrm{D}}}\xspace}
\newcommand{\Mpl}{\ensuremath{{M_\mathrm{Pl}}}\xspace}
\newcommand{\Rinv} {\ensuremath{{R}^{-1}}\xspace}

%
%
\hyphenation{en-viron-men-tal}

\cmsNoteHeader{09-001}
\title{Commissioning and Performance of the CMS Pixel Tracker with Cosmic Ray Muons}

\address[zurich]{University of Zurich - Physik Institut}
		   \address[rice]{Rice University}
\author[zurich]{Vincenzo Chiochia}\author[rice]{Karl Ecklund}

\date{\today}

\abstract{
The pixel detector of the Compact Muon Solenoid experiment consists of three barrel layers and two disks for each endcap. The detector was installed in summer 2008, commissioned with charge injections, and operated in the 3.8~T magnetic field during cosmic ray data taking. This paper reports on the first running experience and presents results on the pixel tracker performance, which are found to be in line with the design specifications of this detector.
The transverse impact parameter resolution measured in a sample
of high momentum muons is 18 microns.
}

\hypersetup{%
pdfauthor={Vincenzo Chiochia,Karl Ecklund},%
pdftitle={Commissioning and Performance of the CMS Pixel Tracker with Cosmic Ray Muons},%
pdfsubject={CMS},%
pdfkeywords={CMS, tracking, pixel, cosmic ray, CRAFT}}

\maketitle 


\section{Introduction}

The Compact Muon Solenoid (CMS) experiment~\cite{:2008zzk} is designed to explore physics at the TeV energy scale exploiting the proton-proton collisions delivered by the Large Hadron Collider (LHC)~\cite{Evans:2008zzb}. The CMS silicon 
tracker~\cite{trackertdr,trackertdraddendum} consists of 1440 silicon pixel and 15\,148 silicon strip detector modules. It is located, together with the electromagnetic and hadron calorimeters, inside a superconducting solenoidal magnet, which provides an axial field of 3.8~T. Outside of the solenoid, the muon system is used both for triggering on muons and for reconstructing their trajectories in the steel of the magnet return yoke.

The pixel tracker allows the reconstruction of charged particle trajectories in the region closest to the interaction point. Installed in July 2008, it is a key component for re\-con\-struc\-ting interaction vertices and displaced vertices from heavy quark decays in an environment characterized by high particle multiplicities and high irradiation.

CMS uses a right-handed coordinate system, with the origin at the nominal interaction point, the $x$-axis pointing to the center of the LHC, the $y$-axis pointing up (perpendicular to the LHC plane), and the $z$-axis along the anticlockwise-beam direction. The polar angle ($\theta$) is measured from the positive $z$-axis and the azimuthal angle ($\phi$) is measured from the positive $x$-axis in the $x$-$y$ plane, whereas the radius ($r$) denotes the distance from the $z$-axis.

\begin{figure}[hbt]
  \begin{center}
    \mbox{
      \subfigure[]
{\scalebox{0.60}{
	  \includegraphics[width=\linewidth]{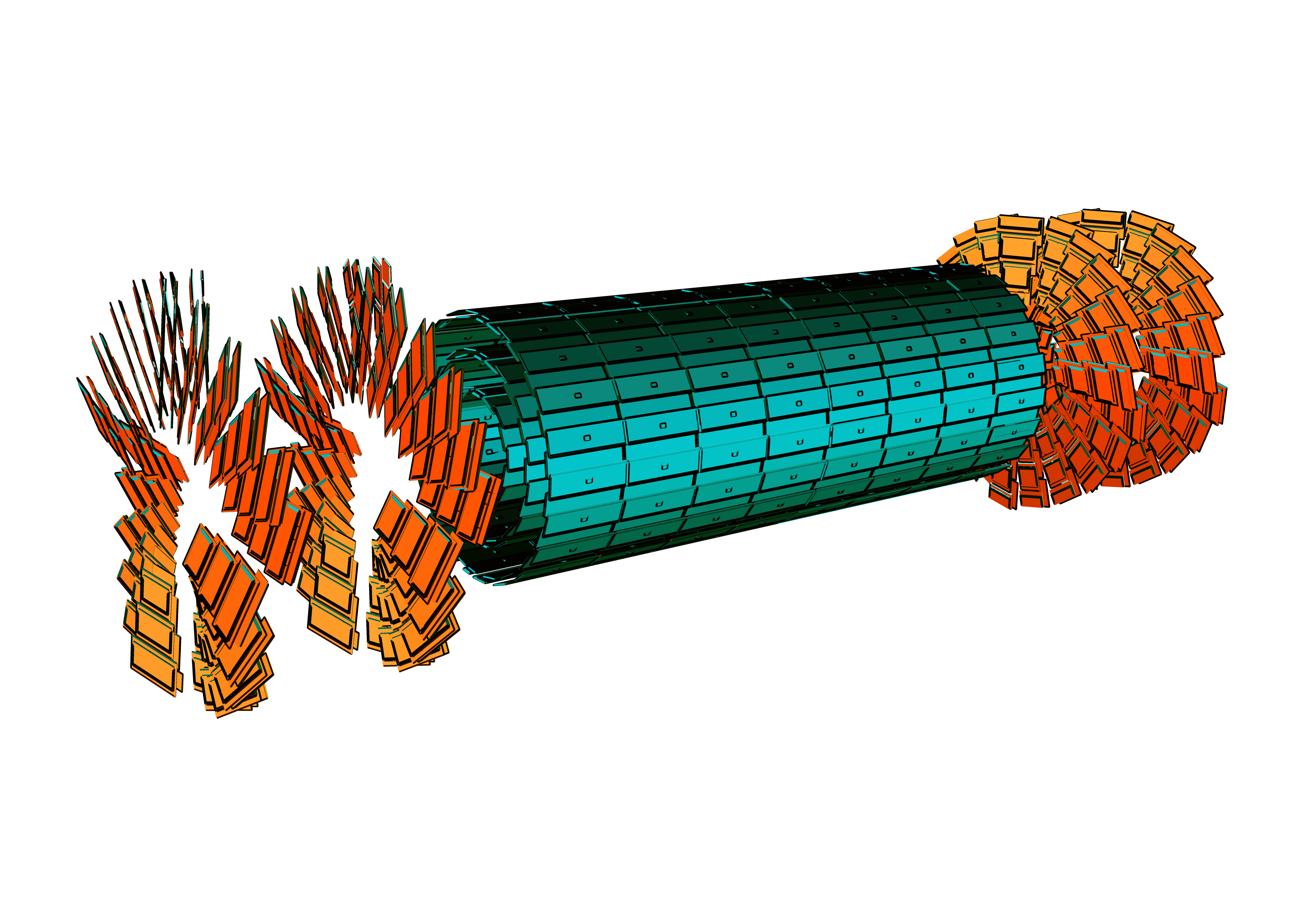}
	  \label{fig:pixeldetector}

      }}
    }
    \hspace{2mm}
    \mbox{
      \subfigure[]
{\scalebox{0.30}{
    \includegraphics[width=\linewidth]{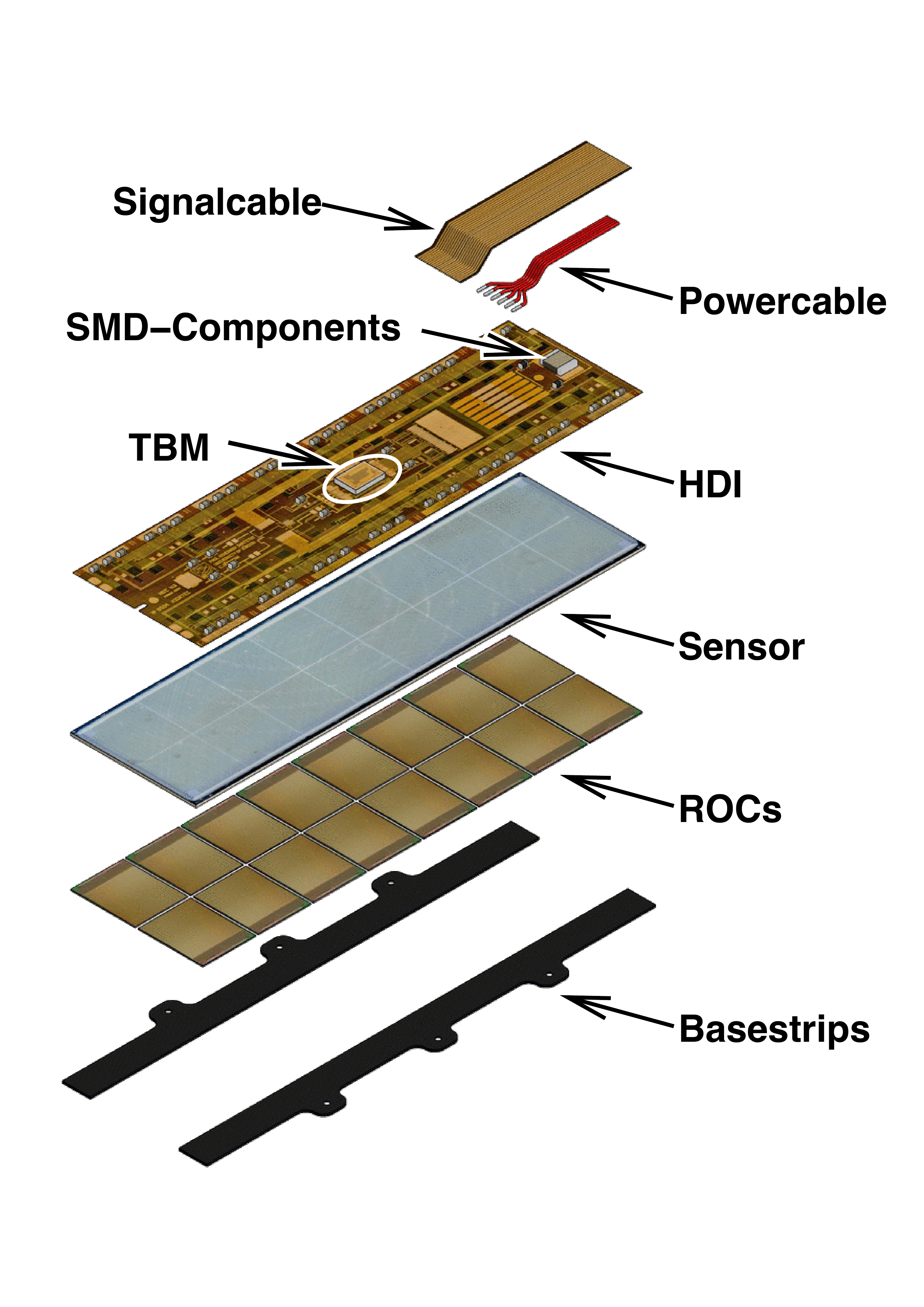}
	  \label{fig:bpixmodule}
      }}
    }

    \caption{Sketch of the CMS pixel detector (a) and exploded view of a barrel module (b).}
  \end{center}
\end{figure}

The pixel tracker consists of three 53.3~cm long barrel layers and two endcap disks on each side of the barrel section, as shown in Fig.~\ref{fig:pixeldetector}. The innermost barrel layer has a radius of 4.4~cm, while for the second and third layers the radii are 7.3~cm and 10.2~cm, respectively. The layers are composed of modular detector units (called {\it modules}) placed on carbon fiber supports (called {\it ladders}). Each ladder includes eight modules, shown in Fig.~\ref{fig:bpixmodule}, consisting of thin, segmented $n$-on-$n$ silicon sensors with highly integrated readout chips (ROC) connected by indium bump-bonds~\cite{Konig:2007pd,Broennimann:2005qv}.
Each ROC~\cite{Kastli:2005jj} serves a 52$\times$80 array of 150 $\mu$m $\times$ 100 $\mu$m pixels.
The ladders are attached to cooling tubes, which are part of the mechanical structure. The barrel region is composed of 672 full modules and 96 half modules, each including 16 and 8 ROCs, respectively. The number of pixels per module is 66\,560 (full modules) or 33\,280 (half modules)~\cite{Amsler:2009sd}. The total number of pixels in the barrel section is 47\,923\,200.

The endcap disks, extending from 6 to 15 cm in radius, are placed at $z=\pm35.5$ cm and $z=\pm48.5$ cm. 
Disks are split into half-disks, each including 12 trapezoidal blades arranged in a turbine-like geometry.
Each blade is a sandwich of two back-to-back panels around a U-shaped cooling channel.  Rectangular sensors of five sizes are bump-bonded~\cite{Merkel:2007zz} to arrays of ROCs, forming the so-called {\it plaquettes}. Three (four) plaquettes are arranged on the front (back) panels with overlap to provide full coverage for charged particles originating from the interaction point.
The endcap disks include 672 plaquettes, for a total of 17\,971\,200 pixels~\cite{Bortoletto:2007zz}.

The minimal pixel cell area is dictated by the readout circuit surface required for each pixel. In localizing secondary decay vertices both transverse ($r\phi$) and longitudinal ($z$) coordinates are important and a nearly square pixel shape is adopted. Since the deposited charge is often shared among several pixels, an analog charge readout is implemented. Charge sharing enables interpolation between pixels, which improves the spatial resolution. In the barrel section the charge sharing in the $r\phi$-direction is largely due to the Lorentz effect. In the endcap pixels the sharing is enhanced by arranging the blades in the turbine-like layout.
The barrel sensors have a sensitive thickness of 285~$\mu$m, and the pixel size is 100~$\mu$m and 150~$\mu$m along the $r\phi$ and $z$ coordinates, respectively~\cite{Allkofer:2007ek}.
The endcap sensors are 270~$\mu$m thick, with the same pixel size oriented in the $r$ (100~$\mu$m) and $r\phi$ (150~$\mu$m) coordinates.
To avoid insensitive areas in-between ROCs, double-sized pixels are located along the three ROC edges not including the chip periphery~\cite{Kastli:2005jj}.

One of the greatest challenges in the design of the pixel detector was the high radiation level expected on all components at very close distances to the colliding beams. At the design LHC luminosity of 10$^{34}$~cm$^{-2}$s$^{-1}$ the innermost barrel la\-yer will be ex\-po\-sed to a yearly par\-ti\-cle flu\-ence
of $3~\times~10^{14}$~n$_{\rm{eq}}$cm$^{-2}$.
Assuming a gradual increase of the LHC luminosity, all components of the pixel system are designed to stay operational up to a particle fluence of at least $6~\times~10^{14}$~n$_{\rm{eq}}$cm$^{-2}$. Test beam measurements have shown that the sensors can survive fluences up to $10^{15}$~n$_{\rm{eq}}$cm$^{-2}$ with breakdown voltages above 600~V~\cite{Allkofer:2007ek,Arndt:2003ck,Bolla:2003si,Cerati:2009zz}.

The pixel tracker performance is expected to evolve with the exposure to irradiation and consequential change of the electric field profile within the silicon bulk. 
The sensor bias voltage will have to be increased with increasing irradiation to compensate for the charge losses due to charge-carrier trapping. Thus, the hit reconstruction software is designed to cope with a varying charge collection efficiency, and to precisely measure the hit position throughout the detector lifetime. The reconstruction techniques rely on periodic calibration procedures~\cite{Chiochia:2008xc,Kotlinski:2009zz}.

This paper describes the detector calibration procedures and reports on early results from data collected with a cosmic ray muon trigger. Detector calibrations are described in Section~\ref{sec:calibrations}. The collected data samples and event selection steps are detailed in Section~\ref{sec:dataset}, and first results from data collected with the cosmic ray muon trigger are presented in Section~\ref{sec:results}.

\section{Detector calibration\label{sec:calibrations}}
This section describes the calibration of the pixel detector in fall
2008 immediately after detector installation and prior to the
cosmic ray run described in Section~\ref{sec:dataset}.  
This first calibration and commissioning included adjustment of the readout
chain, calibration of the pixel charge measurement, and determination of pixel 
readout thresholds.  A summary of hardware problems observed in 2008
is given.

\subsection{Data acquisition electronics}
The pixel data acquisition system is described in more detail in Ref.~\cite{Kotlinski:2006jz}.  A brief description relevant to the detector calibration follows.  The readout chain starts in the pixel cell of the ROC~\cite{Kastli:2005jj}, where the signals from individual pixels are amplified and shaped.  To reduce the data rate, on-detector zero suppression is performed with adjustable thresholds for each pixel.  Only pixels with  charge above threshold are accepted by the ROC, marked with a time-stamp derived from the 40 MHz LHC bunch crossing clock, and stored on chip for the time of the trigger latency (about 3.7 $\mu$s) until readout.  
For each Level-1 trigger, an on-detector ASIC, the Token Bit Manager (TBM), initiates a serial readout of the ROCs on one barrel module or endcap panel.  
In turn, each ROC sends hits matching the trigger bunch crossing in a 40~MHz analog data packet, which encodes pixel address and charge information as described in Section~\ref{sec:ROChain}.
Electrical signals from the TBM are translated by the Analog Optical Hybrid (AOH), and transmitted via optical fiber to off-detector electronics.  

For the full detector, 1214 analog optical links are received in the underground service cavern by forty 36-channel Front End Driver (FED) VME modules.  Each FED has analog optical receivers, flash ADCs, and FPGAs that decode the analog data packets from each channel into pixel addresses and digitized charge information, assemble data packets for each trigger, and buffer the output for transmission of the raw data to the CMS central data acquisition system.  

The ROC includes a charge injection circuit that is used to verify that each pixel cell is functional. It also provides a means to calibrate the ADCs and measure the thresholds by scanning the amount of injected charge.  
Data acquisition and control software runs on eight PCs connected to the three FED VME crates and three other VME crates containing control electronics.  The software performs online calibrations, iteratively adjusting the parameters of the readout chain (Section~\ref{sec:ROChain}), measuring the ADC response to injected charge (Section~\ref{sec:ADCCalibration}), and determining the pixel thresholds (Section~\ref{sec:Thresholds}).
%
%

\subsection{Calibration of the readout chain}
\label{sec:ROChain}
In the 40 MHz analog data packet, six clock cycles are used to encode each hit pixel: the address is amplitude encoded using a six-level scheme over five clock cycles, and the sixth clock cycle gives the pixel 
charge~\cite{Kastli:2005jj}.  A flash ADC on the FED records the data packets, 
and the FPGA firmware decodes the address for each pixel hit.
To operate the 40 MHz analog readout links and properly decode the
analog data packets, a number of online data acquisition calibrations are
performed sequentially to adjust each component of the analog readout chain: ROC and TBM output offsets and gains, 
Analog Optical Hybrid laser bias and gain, FED optoreceiver and channel offsets, 
and FED flash ADC clock delay. 
%
At each stage the signal must remain in the dynamic range of subsequent elements of the readout chain and have sufficient amplitude to be reliably decoded when received at the FED.

\begin{figure}[hbtp]
\begin{center}
    \mbox{
      \subfigure[]
{\scalebox{0.30}{
	  \includegraphics[width=\linewidth, angle=90]{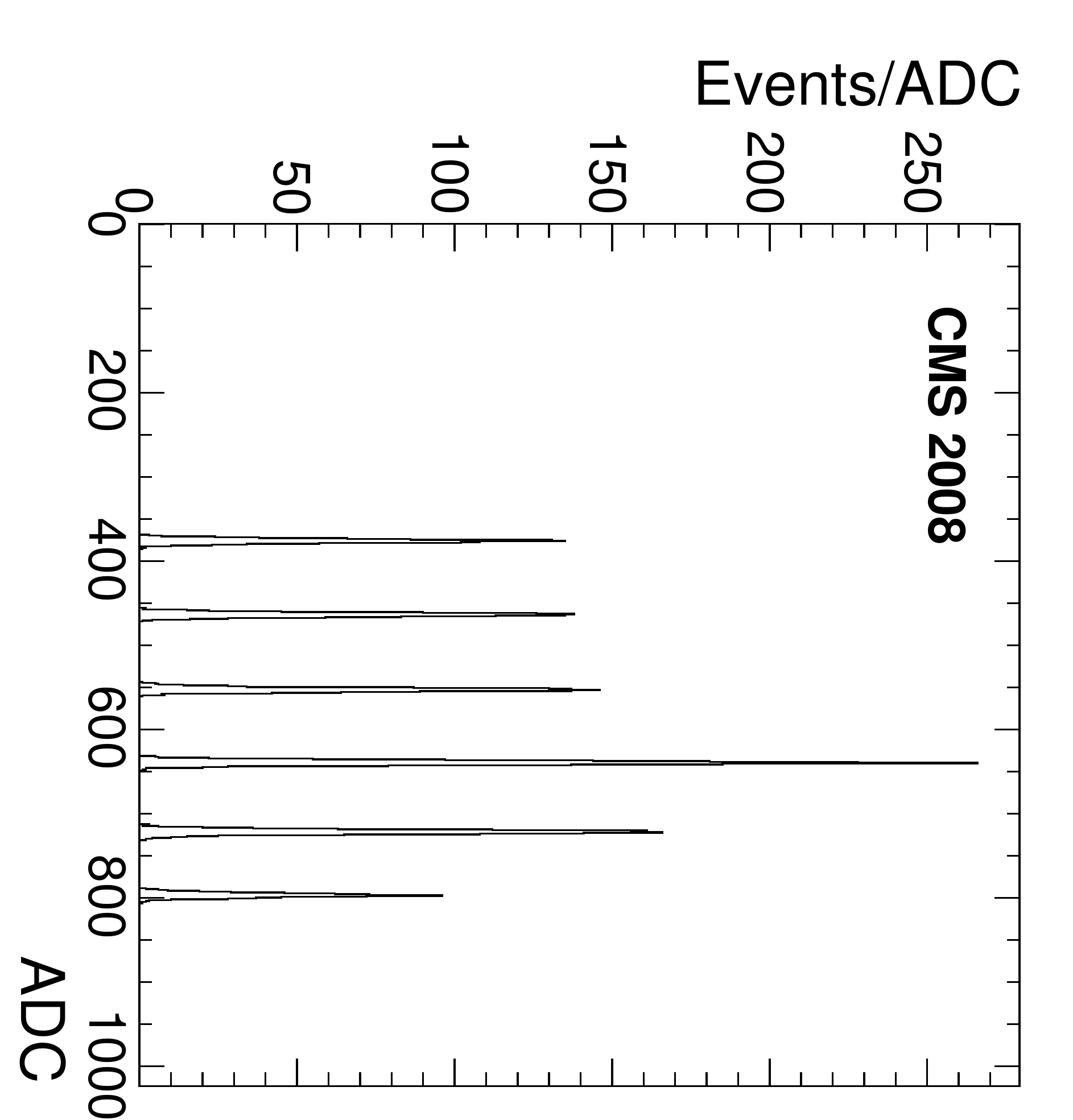}
	  \label{fig:ROClevels} 
      }}
    }
    \mbox{
      \subfigure[]
{\scalebox{0.30}{
	  \includegraphics[width=\linewidth, angle=90]{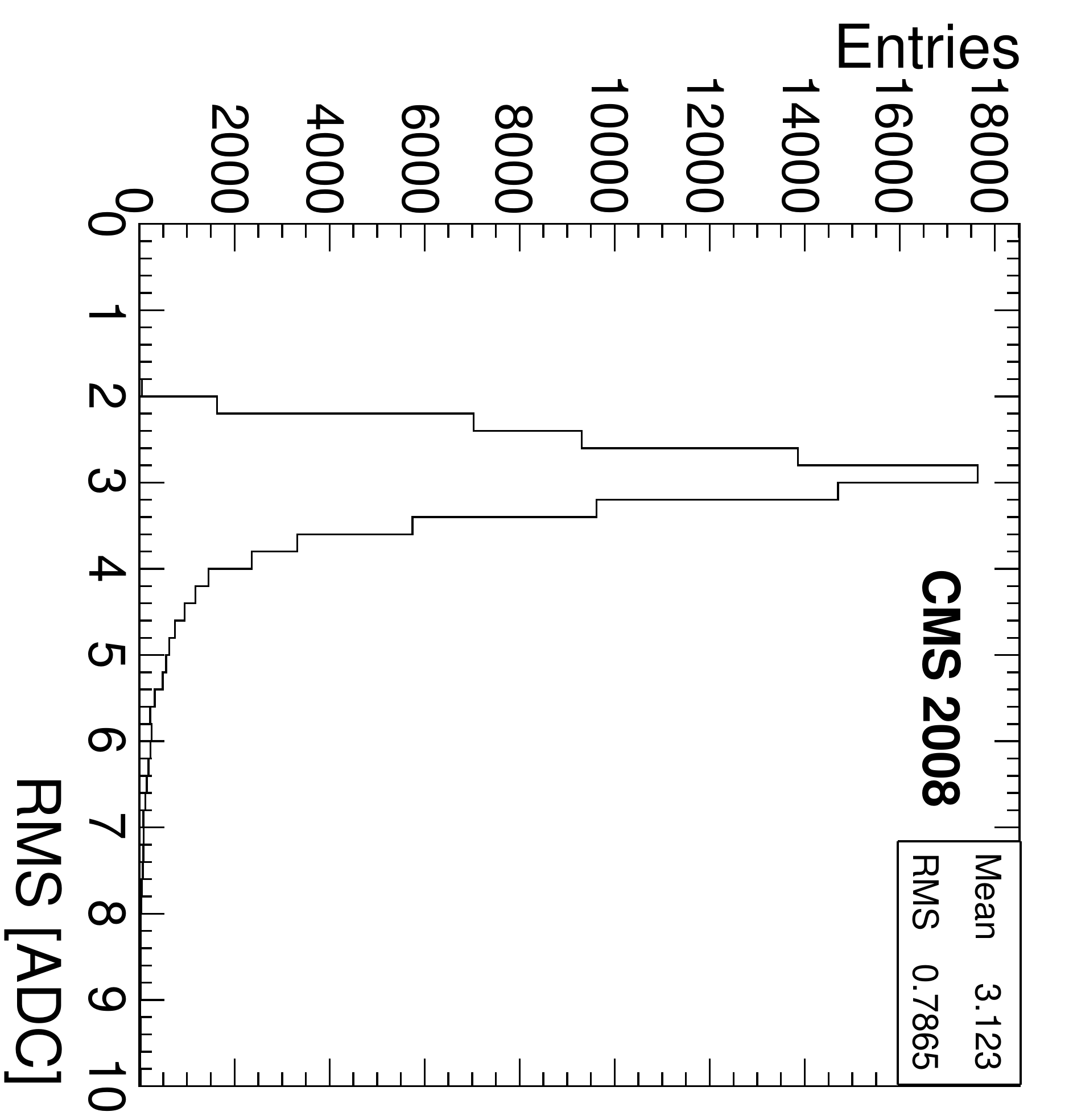}
	  \label{fig:ROClevelsRMS} 
      }}
    }
    \mbox{
      \subfigure[]
{\scalebox{0.30}{
	  \includegraphics[width=\linewidth, angle=90]{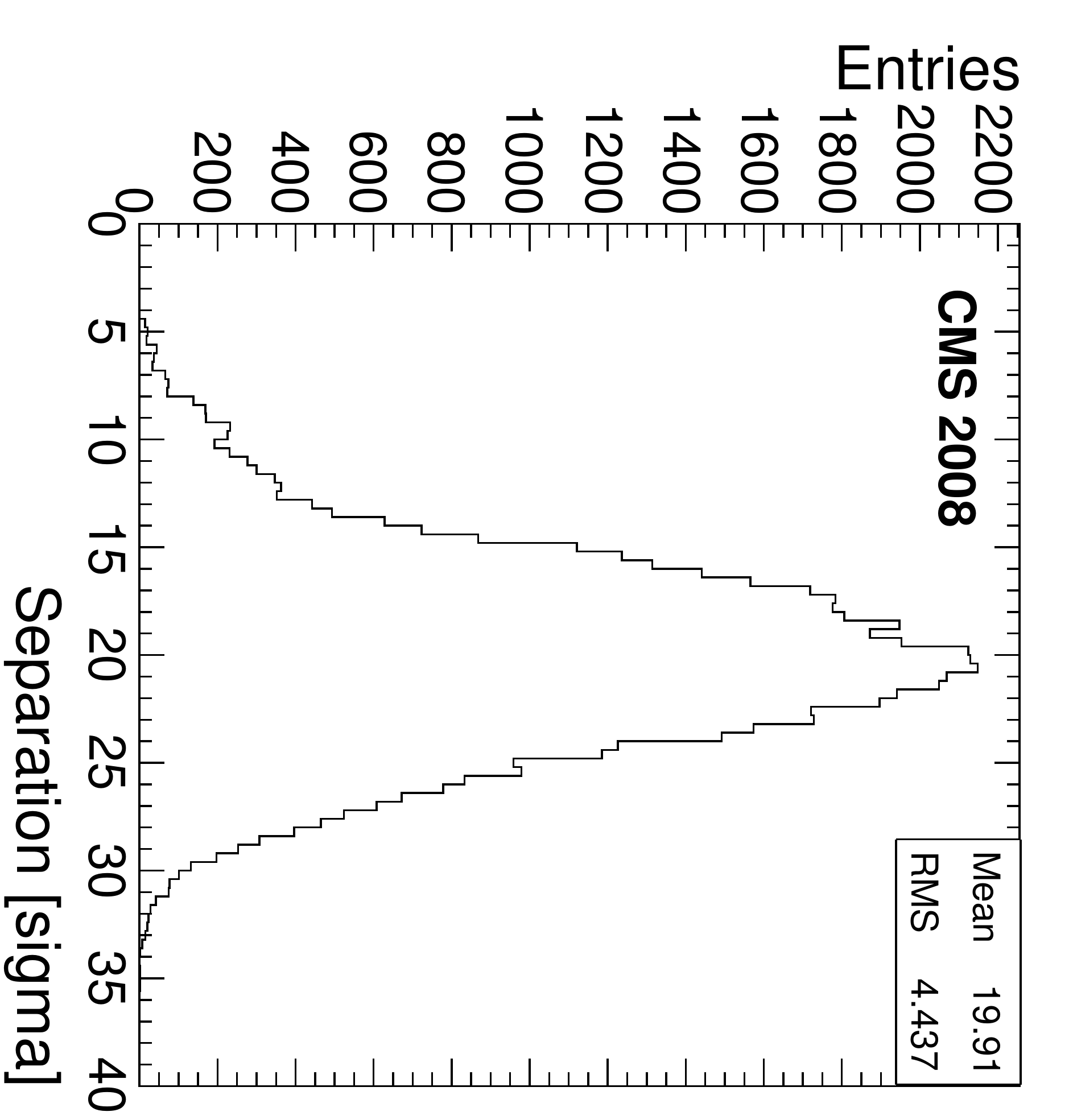}
	  \label{fig:ROClevelsSeparation} 
      }}
    }
\caption{
  (a) ADC values corresponding to
  the address encoding for a single ROC, where six peaks
  corresponding to the levels are visible and well separated. 
  (b) RMS of each peak for all active ROCs in the
  detector.
  (c) Separation between adjacent peaks.}
\label{fig:addresslevels}
\end{center}
\end{figure}

With the amplitude offsets, gains, and timing of the entire readout chain
adjusted, the six ADC levels corresponding to the address encoding are determined from raw ADC information in a dedicated address level calibration
run.  Charge-injection data are collected from each pixel, and all ADC values from the clock cycles corresponding to the address part of the data packet are histogrammed.
A sample set of six address level peaks from one ROC is seen in Fig.~\ref{fig:ROClevels}.  The uneven population of the peaks reflects the choice of encoding for the 4160 pixels on the ROC.
Once determined, the address levels are programmed to the FED FPGA and used to decode pixel data in the FED firmware during subsequent runs.  In rare cases additional adjustments to the front-end parameters or to FED timing are required to achieve good level separation.

To characterize the overall performance of the analog links and
encoding in 2008, the root-mean-square (RMS) widths of all peaks on all
operable readout chips are shown in Fig.~\ref{fig:ROClevelsRMS}. No ROCs were
removed from the analysis. The RMS is typically 3 ADC counts, with the
broadest peaks still less than 7 ADC counts. For a 10-bit ADC, the
level separation is many times the width of the
address level peaks, as seen in Fig.~\ref{fig:ROClevelsSeparation}. Here the separation between the mean of two adjacent address level peaks is given in units of sigma, defined by summing in quadrature the RMS widths for the adjacent peaks.  

\subsection{ADC-to-charge calibration}
\label{sec:ADCCalibration}
Conversion of pixel charge measurements from ADC counts to charge units
requires calibration of the net response function of the pixel readout
chain.  This calibration is essential to achieve a precise hit position,
as the cluster position is interpolated using the charge information from
all pixels in the cluster~\cite{Chiochia:2008xc}.  The most probable charge deposition for normally incident minimum-ionizing tracks is approximately 
21\,000 electrons, as expected for fully depleted 270--285 $\mu$m thick
sensors.  This charge is frequently deposited over more than one pixel
due to Lorentz drift and diffusion of collected electrons.

For each pixel the pulse height response in the 8-bit ADC to a given amount
of collected charge is measured using the charge injection feature of
the ROC.  For each chip, an 8-bit digital to analog converter (DAC), denoted
VCAL, controls the amount of charge injection on each ROC in two
overlapping ranges, which differ by about a factor of seven.
The high and low ranges are cross calibrated, and measurements given
here are converted to low-range VCAL DAC units.
The combination of the two ranges cover the expected 
dynamic range for pixel hits from tracks of varying impact angle and
momentum, from 2000 to 60\,000 electrons.  The charge injection
circuit is approximately linear up to 90\,000 electrons, well beyond
the saturation point of the front-end amplifiers in the readout
chain.  
The calibration of the internally injected charge is obtained using barrel module test data from x-ray sources of known energies \cite{Trueb:2008}.  All barrel modules were tested, finding an average slope of 65.5 electrons per VCAL unit and an offset of $-414$ electrons.  The test data show that the calibration varies by 15\% among the chips.
A pixel-to-pixel variation of similar size is expected, consistent with the typical systematic variation of the charge-injection coupling capacitors in
each pixel cell.  Lacking more detailed test data, the average charge injection calibration is applied to
all pixels in the detector.  The systematic uncertainty on the calibration of the charge injection circuit can be reduced in the future by using isolated tracks in beam collision data.

\begin{figure}[hbtp]
\begin{center}
    \mbox{
      \subfigure[]
{\scalebox{0.30}{
	  \includegraphics[width=\linewidth,angle=90]{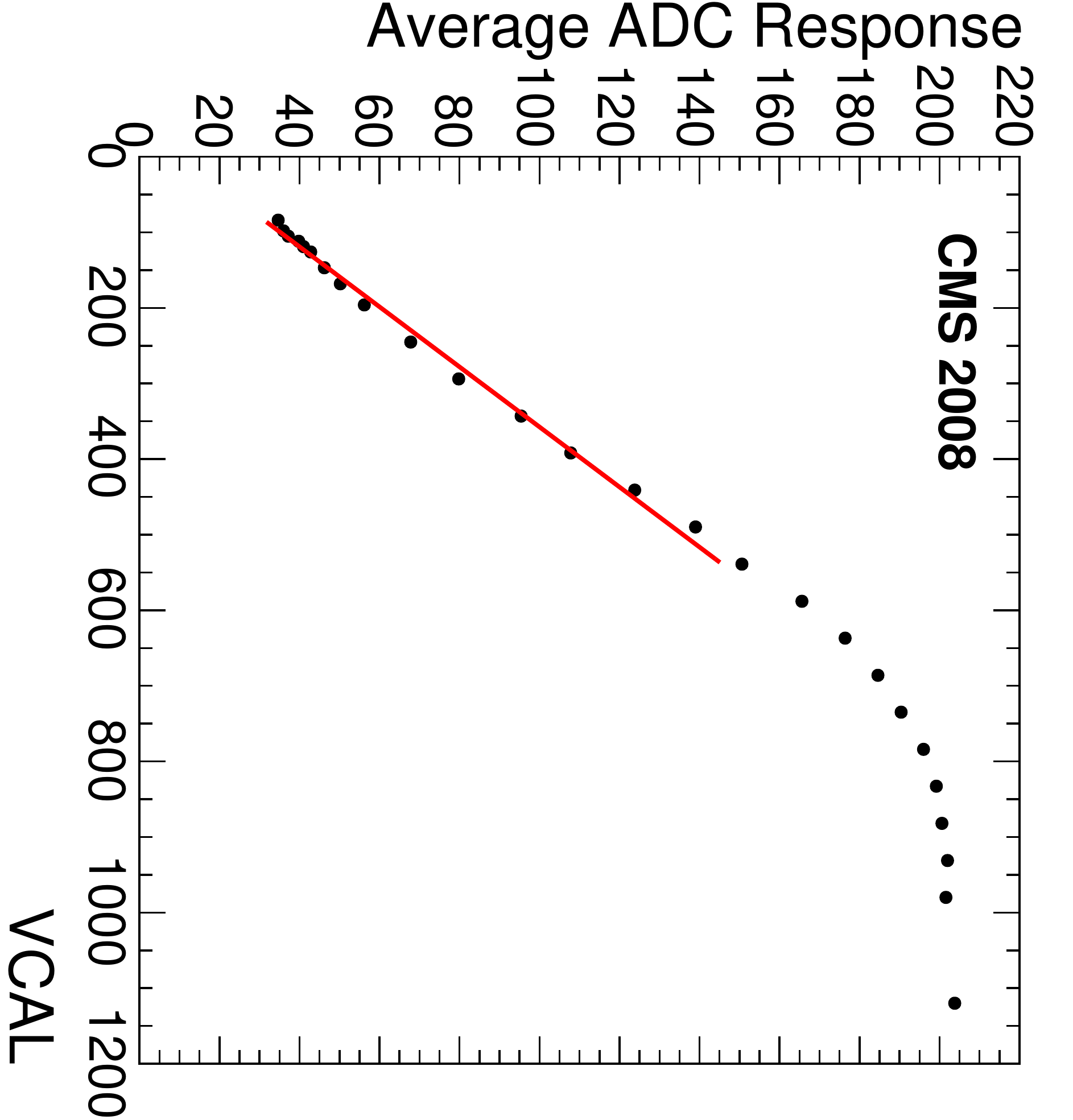}
	  \label{fig:GainCurve} 
      }}
    }
    \mbox{
      \subfigure[]
{\scalebox{0.30}{
	  \includegraphics[width=\linewidth,angle=90]{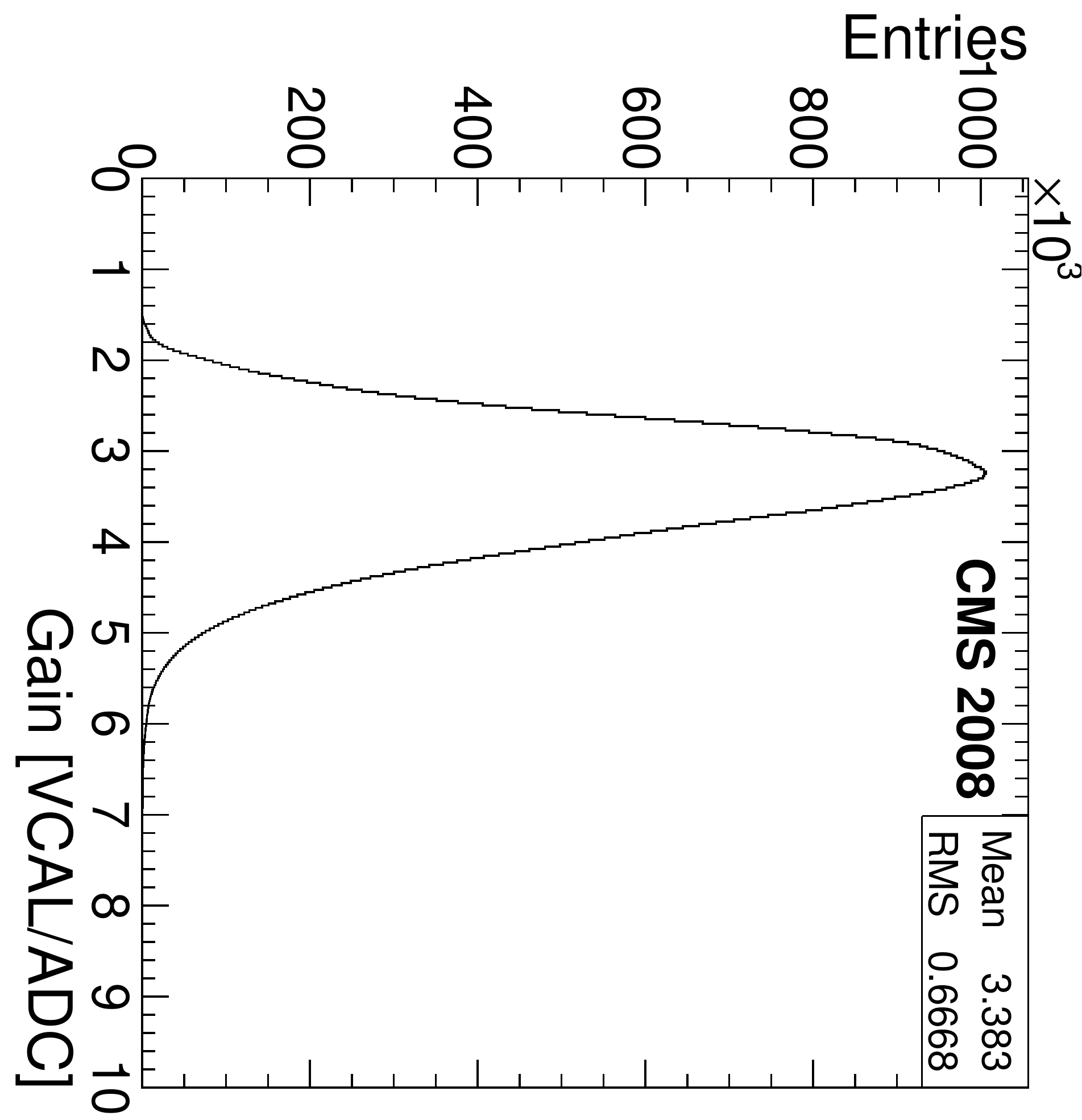}
	  \label{fig:Gains} 
      }}
    }
    \mbox{
      \subfigure[]
{\scalebox{0.30}{
	  \includegraphics[width=\linewidth,angle=90]{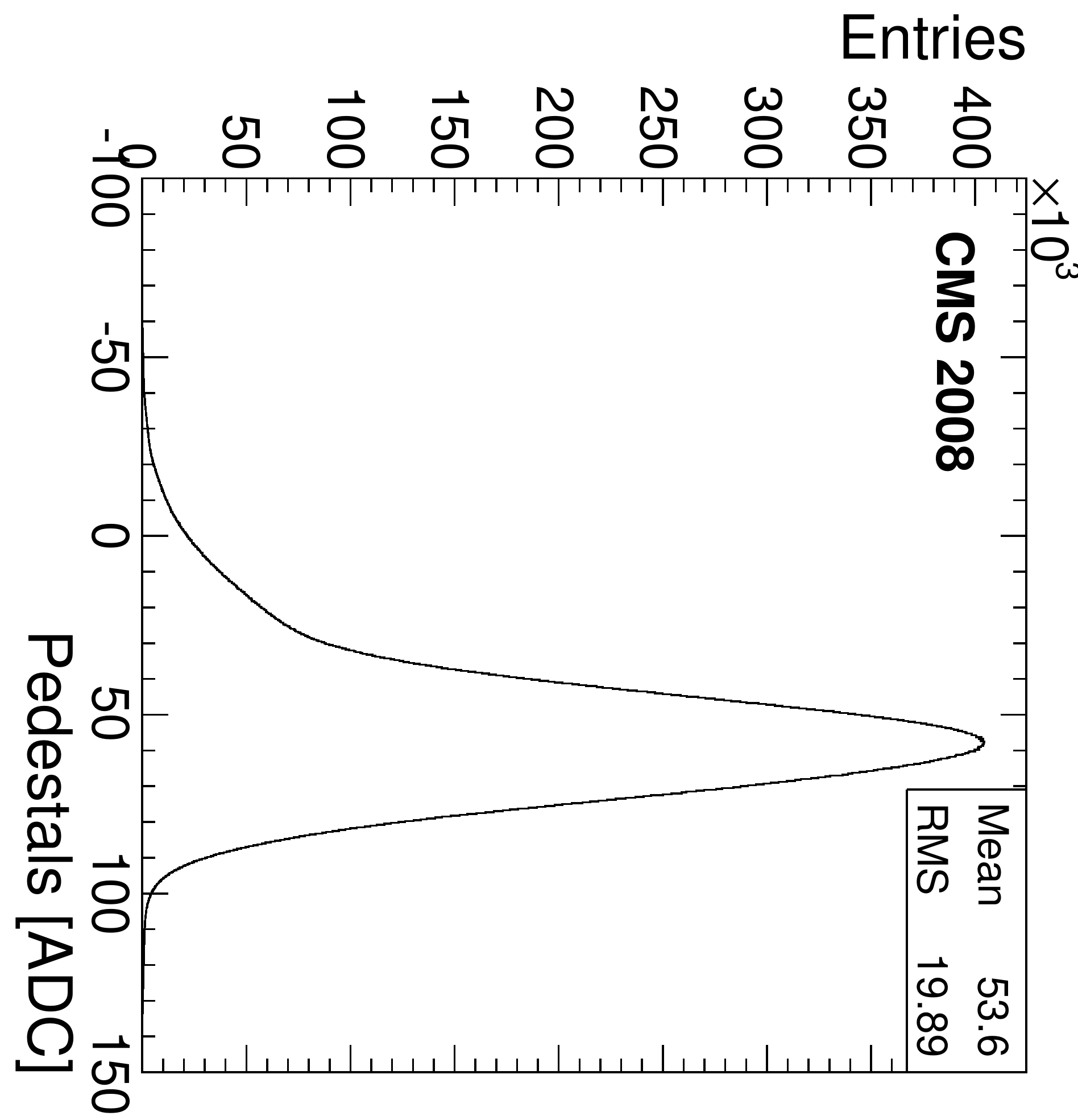}
	  \label{fig:Pedestals} 
      }}
    }
\caption{
  (a) Example of the ADC response as a function of injected charge in VCAL units ($\approx$\,65.5 electrons) for one pixel.
Distribution of gains (b) and pedestals (c) for all pixels.}
\label{fig:GainCalibration}
\end{center}
\end{figure}

Dedicated calibration runs record the ADC response as a function of
the injected charge for all pixels.  Because the pixels will operate
in beam collisions with very low occupancy ($<10^{-4}$), the charge injection
data are also taken with a small fraction of the pixels ($<1\%$)
receiving charge injection on any trigger.  The design of the data
acquisition system demands this low occupancy, but this also minimizes
any effect of crosstalk by geographically separating the
pixels with injected charge.  The pattern of injected pixels is changed to serially cover the entire chip.  For each pattern the injected charge is varied.
The response is approximately linear below saturation at about
45\,000 electrons, and the data are fit with a first degree polynomial in the linear region. An example fit for one pixel is shown in Fig.~\ref{fig:GainCurve}.  

The distributions of the two fit parameters for all pixels are given in
Figs.~\ref{fig:Gains} and \ref{fig:Pedestals}.  The pedestal and gain
parameters correspond to the intercept and the inverse slope,
respectively, in Fig.~\ref{fig:GainCurve}.  The tail in the pedestal
distribution is due to poorly optimized front-end parameters used in some ROCs.  
Since fall 2008, the ROC parameters have been adjusted to remove the negative pedestals.

\subsection{Readout thresholds}
\label{sec:Thresholds}
The thresholds are controlled at pixel level using two 8-bit DACs
per ROC plus one 4-bit trim value per pixel.  The pixel threshold is
given by a global chip threshold DAC, adjusted with the 4-bit trim
within a range set by the second DAC.  The trim bits are used to equalize the thresholds for all pixels on a readout chip.

During the cosmic ray run, conservatively high thresholds were
chosen to ensure stable and efficient operation of the data
acquisition, avoiding inefficiencies from overflowing hit buffers on the
ROCs.  As thresholds decrease, hit rates increase and do so very sharply when reaching the level where internal ROC readout induces crosstalk.  
The minimum threshold is set to avoid crosstalk in the most sensitive pixel  on each chip.
The threshold parameters for the barrel detector were set to
values determined during module testing at construction time.  The
endcap disk thresholds were tuned \textit{in situ} by adjusting the
threshold DACs and trim bits.  

The threshold tuning process is iterative and time consuming.
At each step, the threshold DACs or trim bits are adjusted for a
target threshold, and the threshold is measured, as described below, for a sample of pixels on each ROC, iterating to converge to the target. 
Then all pixels are checked to confirm that  
there is no buffer-overflow inefficiency at the target threshold.
If successful, the thresholds are lowered another step of approximately 300 electrons, and the procedure is repeated.

\begin{figure}[hbtp]
\begin{center}
    \mbox{
      \subfigure[]
{\scalebox{0.45}{
	  \includegraphics[angle=90,width=\linewidth]{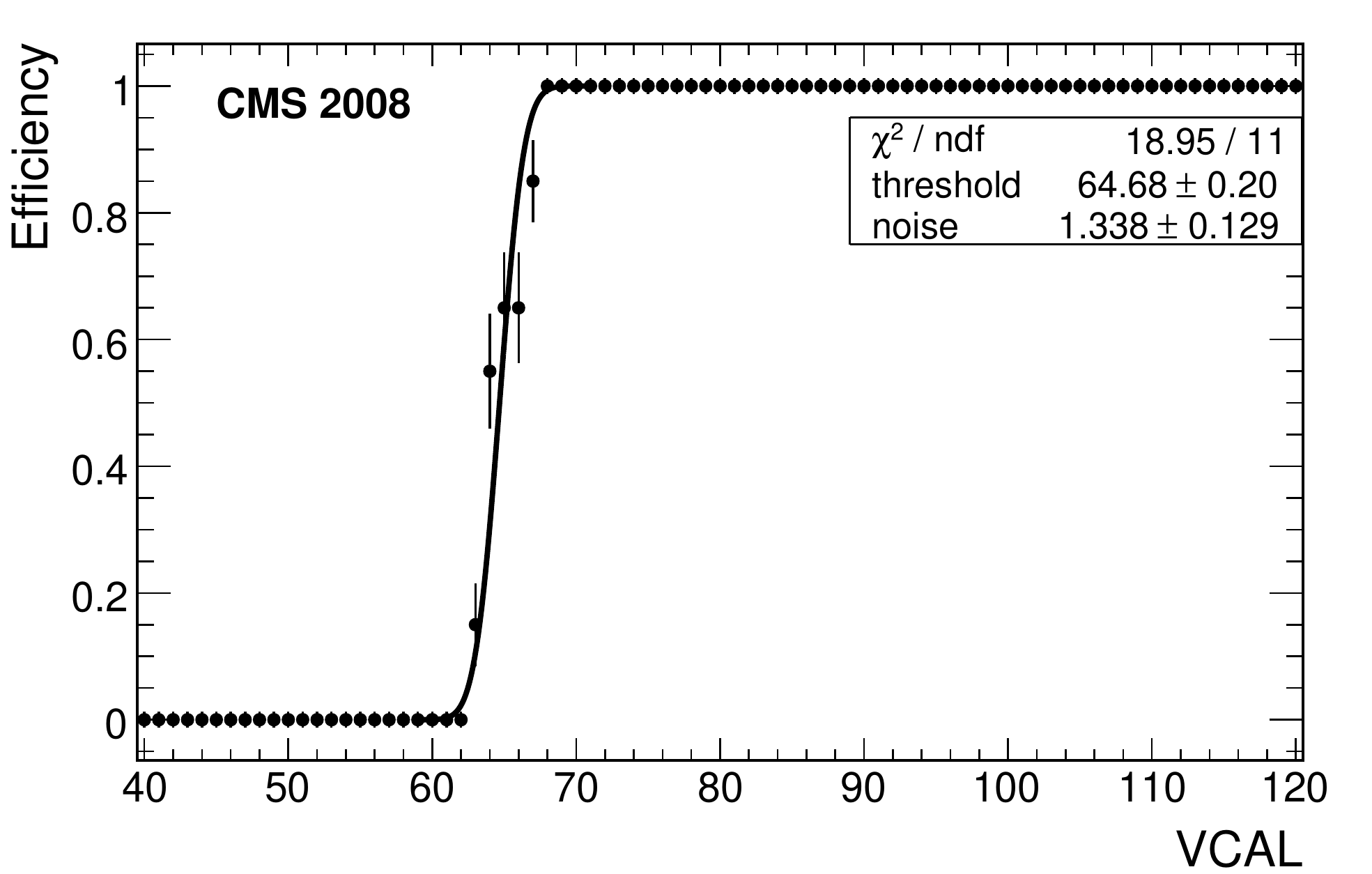}
	  \label{fig:SCurve} 
      }}
    }\hfill
    \mbox{
      \subfigure[]
{\scalebox{0.45}{
\includegraphics[angle=90,width=\linewidth]{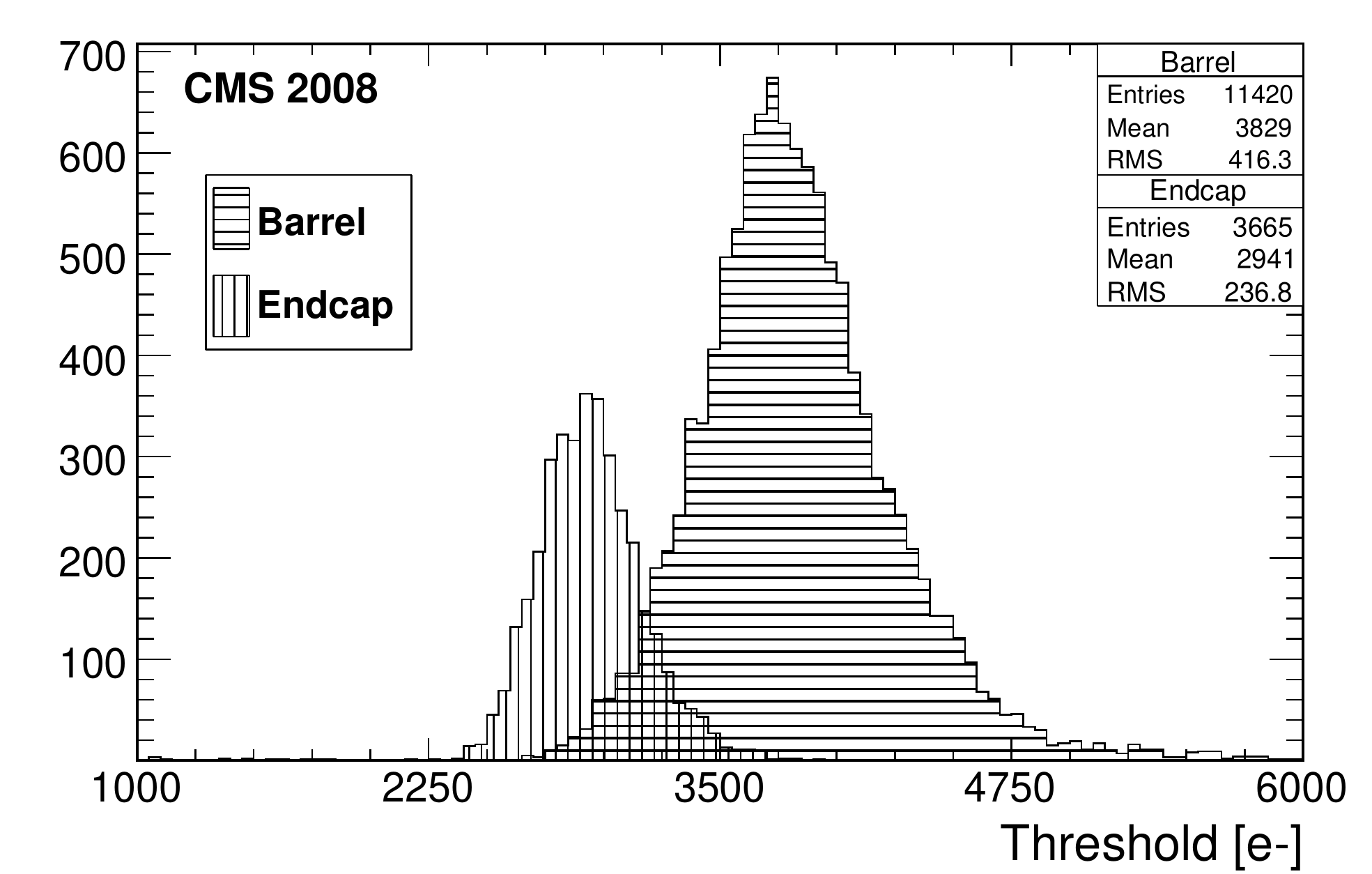}
	  \label{fig:Thresholds} 
      }}
    }

\caption{
  (a) Efficiency S-curve as a function of injected charge in VCAL units ($\approx$65.5 electrons).
  (b) Distribution of ROC-mean threshold in the endcap and barrel detectors.}
\label{fig:threshold}
\end{center}
\end{figure}
The readout thresholds are measured using charge injection runs, performed by the DAQ system in the same way as described for the ADC-to-charge calibration.  For each pixel, the efficiency as a function of injected charge (S-curve) is fit to an error function.  
The efficiency is defined as the number of charge injection events with a pixel hit divided by the number of charge injection events.
The threshold is the injected charge where the efficiency is 50\%.
The noise is measured from the slope of the turn-on region.
The error function fit parameters
are the threshold and noise, defined as the mean and RMS width, respectively, of the Gaussian that gives the error function when integrated.
Figure~\ref{fig:SCurve} shows an example S-curve and fit for one pixel.

Figure~\ref{fig:Thresholds} shows the distribution of the mean threshold per ROC in the barrel and endcap detectors for the fall 2008 cosmic trigger run.  
The overall average thresholds are found to be 3829 and 2941 electrons for the barrel and endcap, respectively.  
Since there was more time available to tune the endcap disks, they
were operated at lower thresholds than the barrel layers.  After the
cosmic run ended, a carefully tuned subset of the detector achieved
lower and more uniform thresholds.  This tuning was extended to the
entire detector in 2009 during a two-week commissioning period.

The thresholds reported above are absolute thresholds, 
valid for hits assigned to any bunch crossing in the ROC.  
Because only a single bunch crossing may be read from the ROC, to
measure the absolute thresholds, the charge injection scan repeats
three times, varying the ROC latency over three consecutive bunch
crossings, and the efficiencies defined above are summed.  
The in-time
threshold, which requires the time-stamp to match the bunch crossing
of charge deposition, turns out to be 600--1000 electrons higher.  The higher
in-time threshold comes from the time-walk effect of the ROC
comparator: due to the finite rise time of the shaped signal in the
ROC, small signals cross the threshold later than large  
signals~\cite{Kastli:2005jj} 
and can be time-stamped in the subsequent
bunch crossing.  Such hits are suppressed at the ROC, which only reads out hits matching the trigger time-stamp.
The in-time threshold is appropriate when considering hit
efficiency, where a hit must be found in-time with the trigger bunch crossing to be transmitted for readout.  The absolute threshold controls the occupancy of on-chip hit buffers and stability of the ROC. See
Ref.~\cite{Kotlinski:2009zz} for additional discussion of pixel thresholds.  The difference in
absolute and in-time thresholds may be reduced by increasing the
bandwidth of the front-end amplifiers, which was not optimized during
the initial commissioning period.  
%

From the threshold measurement with the fall 2008 tuning, the mean noise is 141 and 85 electrons for the barrel and endcap, respectively.
The noise performance of the endcap
detector is expected due to a different
design of the sensors\cite{Allkofer:2007ek,Arndt:2003ck} resulting in
smaller pixel capacitance.
These noise levels are well below operating thresholds, which must be set
above crosstalk levels.  It should be noted that this noise
represents the contribution from the front-end ROC only. The
remainder of the readout chain (amplifiers, laser, and optical
receiver) contributes an additional 300 electrons of noise to the
charge measurement, but has no effect on the S-curve or pixel occupancies.

\subsection{Inactive and noisy channels}
During the run with cosmic ray muon triggers, 99\% of the barrel pixels and
94\% of the endcap pixels were operational.  The remaining
channels were inactive due to a small number of failures, listed in Table~\ref{tab:dead} and discussed in the following.

\begin{table}[htbp]
\caption{\label{tab:dead} Summary of inoperative channels during the fall
  2008 cosmic ray run.}
\begin{center}
\begin{minipage}{0.45\textwidth}
\begin{tabular}{lrr}
\hline
\multicolumn{3}{c}{Barrel Pixels} \\ 
 Cause & \# ROCs & Fraction \\
\hline
No HV              &  40 & 0.35\% \\
Readout wire bond  &  40 & 0.35\% \\
Dead module        &  16 & 0.14\% \\
Bad ROC            &   4 & 0.03\% \\ 
\hline
Total              & 100 & 0.87\% \\
\hline
\end{tabular}
\end{minipage}
\hfill
\begin{minipage}{0.45\textwidth}
\begin{tabular}{lrr}
\hline
\multicolumn{3}{c}{Endcap Pixels}\\ 
 Cause & \# ROCs & Fraction \\
\hline
Shorted LV cable & 135 & 3.13\% \\
Shorted HV cable &  93 & 2.15\% \\
Wire bond (HV)   &   8 & 0.19\% \\
Bad TBM header   &  24 & 0.56\% \\ 
\hline
Total            & 260 & 6.02\% \\ 
\hline
\end{tabular}
\end{minipage}
\end{center}
\end{table}

The barrel pixel detector had 100 ROCs that could not be operated, 80 of
which, spread over eight modules, were due to broken wire bonds or missing high voltage connections.
One module (16 ROCs) did not respond to
programming and was disabled.  Four additional ROCs, randomly
scattered throughout the detector, did not respond or produce signals.
The number of dead pixels on otherwise functional ROCs was very low,
0.01\%, and consistent with the fraction of faulty bump bond
connections between the ROC and sensor observed during module testing.
The net inefficiency due to defective connections or hardware was less
than 1\%.  Note that the dynamic inefficiency from overflowing buffers in the readout chain will be a few percent at design luminosity
\cite{:2008zzk,Kastli:2005jj}.

In the endcap pixel detector, a shorted power supply cable resulted
in a loss of six panels (out of 192), 3.13\% of the detector.
Another shorted high voltage cable disabled the two outer plaquettes on
each of six panels (2.15\%).  These two faulty cables accounted for most of
the lost channels in the endcap pixel detector.  They were
repaired in 2009, demonstrating the quick removal, repair, and
reinstallation feature of the CMS pixel design.  An additional plaquette (0.19\%) was insensitive due to a broken wire bond,
detected by naked eye during installation in August 2008.  The
affected panel was also replaced during maintenance.  One other panel
(0.56\%) has an intermittent and temperature dependent connection.  
For the endcaps, dead pixels and faulty bump bonds, measured during construction~\cite{Merkel:2007zz}, add less than 0.1\% to the losses in Table~\ref{tab:dead}.

The number of noisy pixels is negligible.  During the cosmic ray run, a total of 263 barrel and 17 endcap pixels produced hits at a rate of more than
10$^{-3}$ per trigger and were disabled during early running.  
Given the exceptionally low occupancy in cosmic ray events,
a hit rate at this level is clearly due to malfunction or poorly adjusted thresholds.  Changing the criterion for a noisy pixel to a hit rate of
$10^{-4}$ per trigger adds only 8 barrel and 5 endcap pixels.  The
total fraction of noisy pixels was less than $5\times 10^{-6}$.

\section{Data samples, alignment, event selection and simulation\label{sec:dataset}}

The CMS collaboration conducted a month-long data taking exercise known as the Cosmic Run At Four Tesla (CRAFT) during October--November 2008, with the goal of commissioning the experiment for extended data taking~\cite{CRAFTpaper}. 
CMS recorded 270 million cosmic-ray-triggered events with the solenoid at its nominal axial field strength of 3.8~T and the tracking detectors operational. A few percent of those events had cosmic ray muons traversing the tracker volume. Prior to CRAFT and during the final installation phase of the experiment from May to September 2008, a series of commissioning exercises to record cosmic ray events took place with the solenoid turned off.
The reverse bias voltage of the pixel sensors was set to 100~V and
300~V in the barrel and endcap sections, respectively. A higher bias
voltage in the endcap section was used to reduce noise in a limited
number of plaquettes.  These noisy plaquettes were close to the wafer edge during manufacture and require a higher bias to meet performance specifications.


From the data taken with the 3.8 T field, approximately 85\,000 tracks traversing the pixel detector volume were reconstructed with the Combinatorial Track Finder (CTF) algorithm~\cite{tracking}. The algorithm combines hits in the pixel and strip tracker. For this set of tracks, the average number of pixel hits is 3, for a total of about 257\,000 clusters reconstructed in the pixel system. 
All results shown in this paper were obtained using tracks reconstructed by the CTF algorithm. 

Spatial alignment of the pixel detector is detailed in Ref.~\cite{alignment}. The precision of the detector position with respect to particle trajectories after track-based alignment has been derived from the distribution of the median of the cosmic muon track residuals measured in each module. The barrel precision is 3~$\mu$m RMS in the $r\phi$ coordinate and 4 $\mu$m in the $z$ coordinate, while the endcap precision is 14~$\mu$m RMS along both the $r$ and $\phi$ coordinates.

The timing alignment of the pixel modules was performed during the
cosmic data taking.  There are approximately 100 separate optical
fibers distributing the beam crossing clock to the pixel detector,
with each fiber serving a group of modules in close proximity.  The
lengths of these fibers are measured using reflectometry, and a
programmable delay chip (Delay25) is used to equalize the
delay for each of the fiber links.  
The Delay25 chip can be set in increments of 500 ps; after
correction for measured fiber lengths, the total delay is estimated to
be controlled to better than 2 ns.  The inter-pixel timing adjustment
was done \textit{a priori} using fiber length measurements.

With each of the modules aligned to the same delay, a single overall
delay remains to be determined for the pixel detector relative to the
trigger and the rest of CMS.  Starting from an \textit{a priori}
calculation of the total latency in the trigger and fiber delays, a
coarse scan of the pixel detector latency was made in four steps of 25
ns, the period of the beam crossing clock.  Tracks were
reconstructed in the silicon strip tracker, which has wider time
acceptance than the pixel detector. Tracks that cross the cylinder
defining the pixel fiducial volume are used to measure the efficiency
of pixel hits for each delay setting. Two of the beam crossings had
measurable efficiency. Initially the more efficient of the two was
used for about 50\% of the CRAFT data.  Part way through the run the
phase of the beam crossing clock was shifted by 9 ns to maximize the
pixel efficiency.

There are limitations to the ability to make a timing alignment for
cosmic ray tracks.  The cosmic tracks arrive at random phases of the
LHC beam crossing clock, and have a broad distribution in time of
flight from the various muon stations to the pixel detector.  These
two effects give rise to a time distribution that is much wider than
the sharp distribution from beam collisions.  Combined with the single
bunch crossing sensitivity of the readout chip, some inefficiency is
expected in cosmic ray running.  These effects are not expected in
beam collision running.  Information on the timing of cosmic ray
events from the muon detectors is used to minimize these effects in
the analyses that follow.  The muon time measurements have a
mean resolution of approximately 6 ns.  They are corrected for time of
flight to the pixel detector.

The events analyzed in Section~\ref{sec:results} were selected according to the following criteria:
\begin{itemize}
\item Events belonging to runs with stable magnetic field at 3.8~T;
\item Events belonging to runs in which all pixel detector FEDs are included in the data acquisition; 
\item Events with two muon legs reconstructed by the muon detectors (see Section~\ref{sec:trackparresolution}); 
\item The weighted mean of the muons arrival time at the pixel
  detector, as determined by the muon detectors, is required to be
  within $\pm$20 ns with respect to the Level 1 trigger signal, in
  phase with the LHC bunch crossing clock; 
\item The averaged uncertainty on the muon time measurement is required to be smaller than 10 ns.
\end{itemize}

For comparison with the data, event samples generated with {\tt CMSCGEN}~\cite{Biallass:2007zz} are
processed with the {\tt CMSSW} software and include the full detector
simulation~\cite{Bayatian:2006zz}.  We also use the {\tt PIXELAV}
simulation~\cite{Chiochia:2004qh,Swartz:2005vp}, which is a detailed
treatment of the pixel sensor, for the Lorentz angle and resolutions
studies presented in Section~\ref{sec:results}.

\section{Results\label{sec:results}}
\subsection{Hit distributions and charge collection}
\begin{figure}[hbt!]
  \begin{center}
    \mbox{
      \subfigure[]
{\scalebox{0.325}{
	  \includegraphics[angle=90,width=\linewidth]{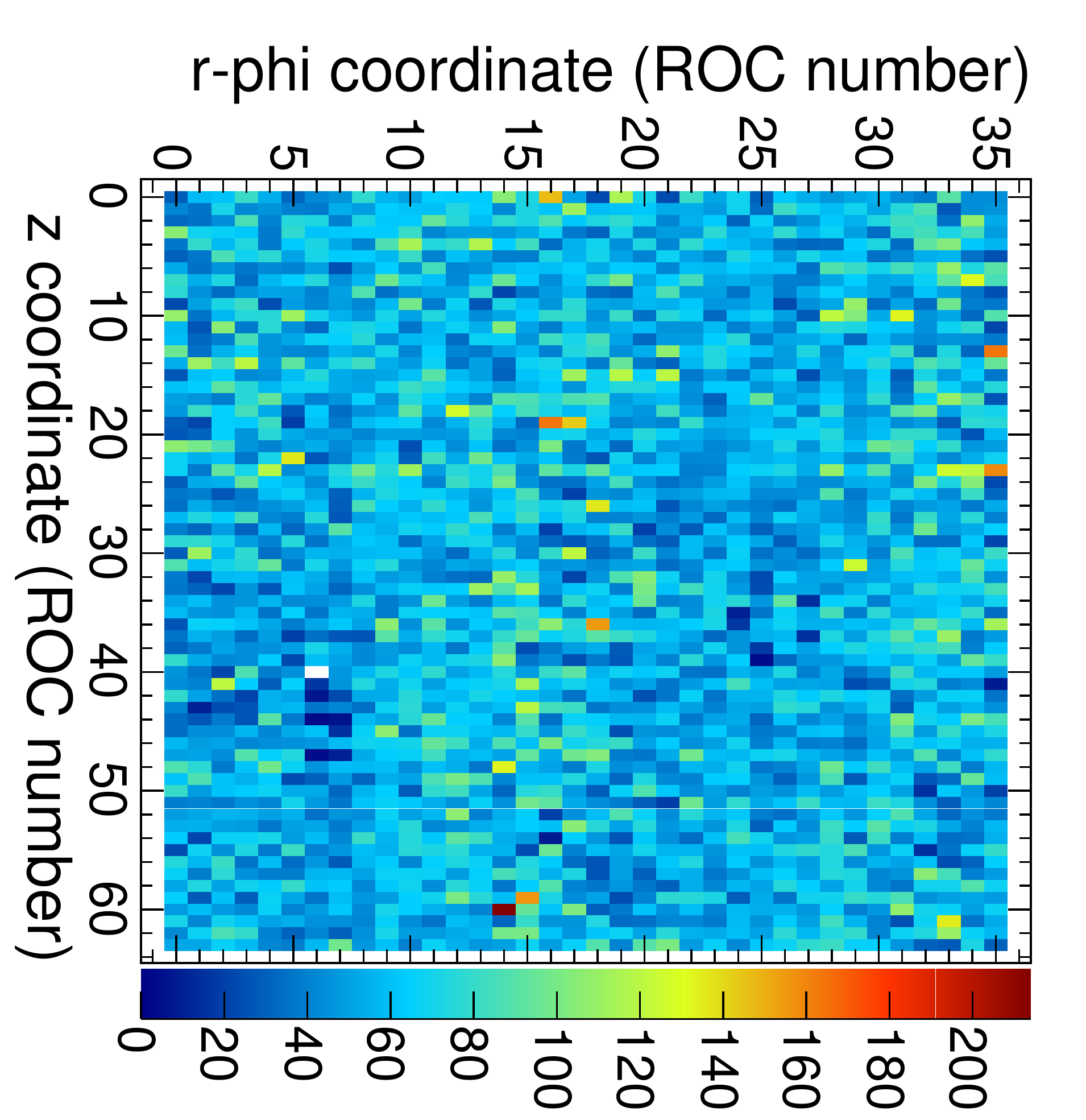}
	  \label{fig:ROC_lay1} 
      }}
    }
    \hspace{-0.5cm}
    \mbox{
      \subfigure[]
{\scalebox{0.325}{
	  \includegraphics[angle=90,width=\linewidth]{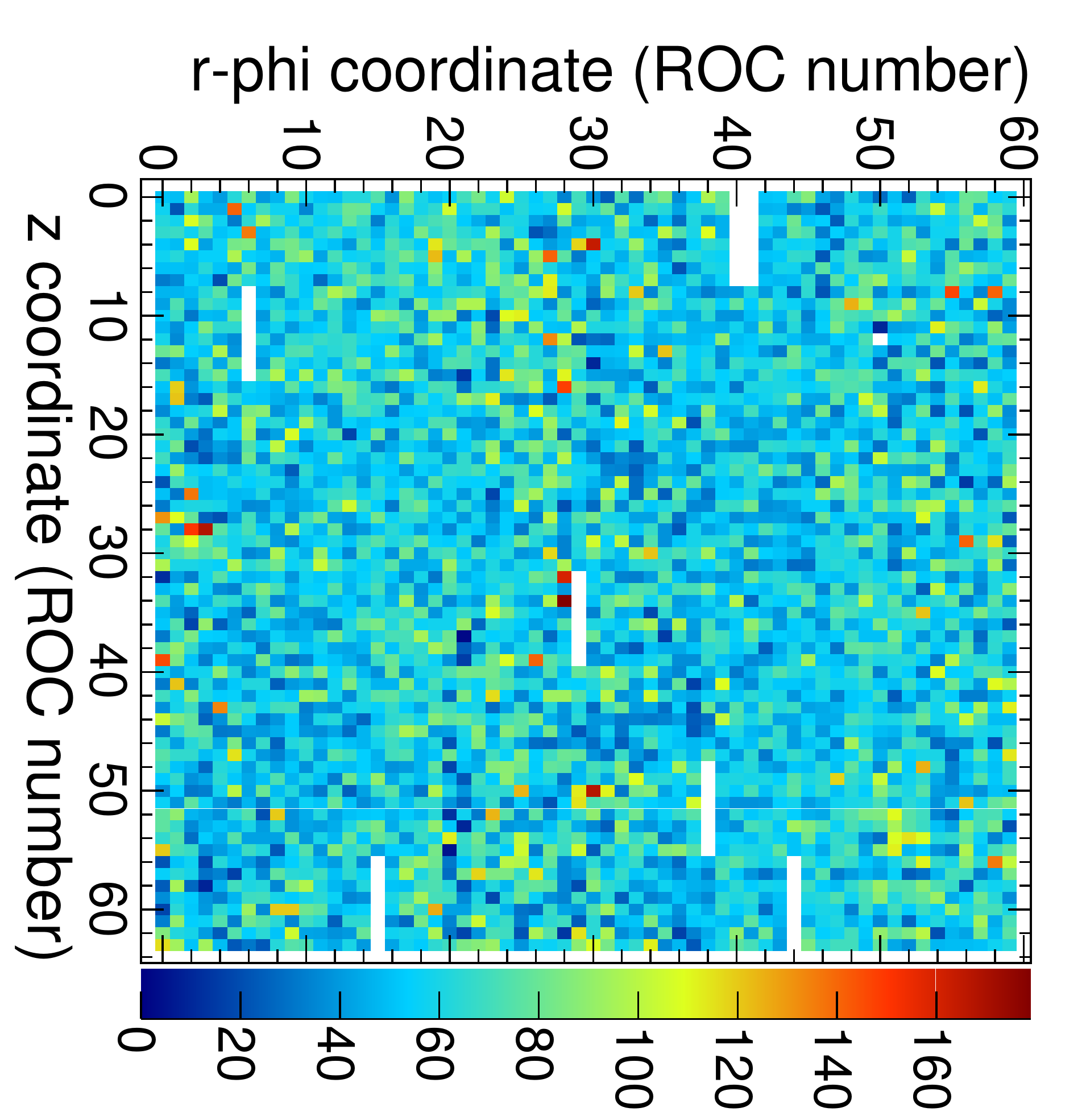}
	  \label{fig:ROC_lay2} 
      }}
    }
     \hspace{-0.5cm}
    \mbox{
      \subfigure[]
{\scalebox{0.325}{
	  \includegraphics[angle=90,width=\linewidth]{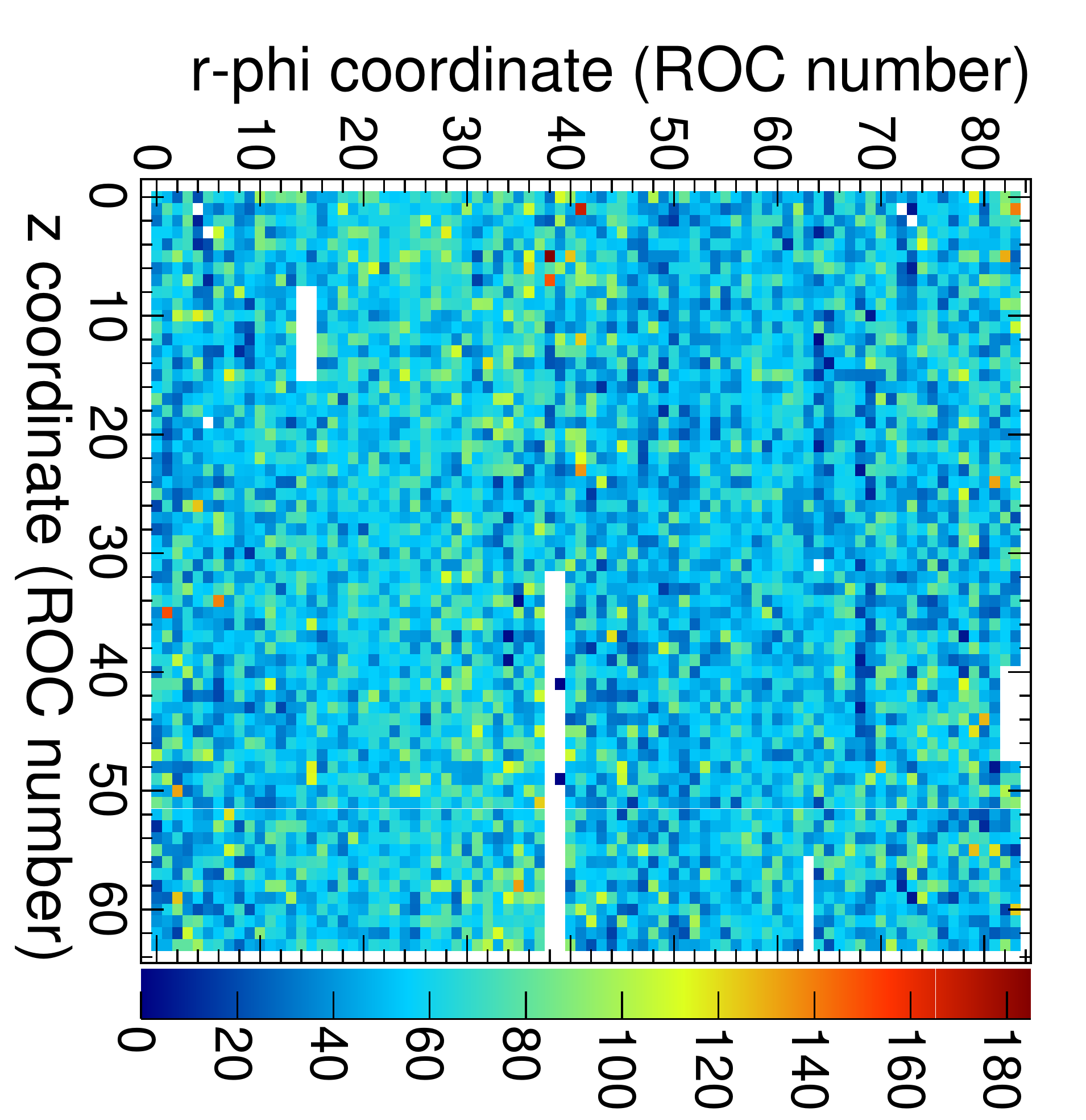}
	  \label{fig:ROC_lay3} 
      }}
    }

    \caption{Number of hits associated to a track detected in each ROC for the first (a), second (b) and third (c) barrel layers. Bins in white correspond to readout chips excluded from data taking. On average each ROC had about 60 hits when integrated over the 85\,000 tracks traversing the barrel pixel detector. The plot origin corresponds to $\phi=0$ and $z=-26.7$ cm.}    \label{fig:hit_occupancy}
  \end{center}
\end{figure}

Pixel clusters are formed from adjacent pixels with a charge above the readout threshold. Both side and corner adjacent pixels are included in the cluster. The charge collected in each pixel is converted into electrons using the calibration procedure described in Section~\ref{sec:calibrations} and cluster projections along the sensitive coordinates are obtained by summing the charge collected in the pixels with the same coordinate. Residual charge miscalibration due to the pixel-to-pixel variation of the charge injection capacitor are extracted from laboratory measurements and included in the Monte Carlo simulation. Clusters with total charge above 5000 electrons are selected and their position is calculated as described in Ref.~\cite{Chiochia:2008xc}.
Figure~\ref{fig:hit_occupancy} shows the number of track-associated hits in each barrel ROC. ROCs showing no hits were excluded from the readout either because of DAQ or sensor biasing problems. 

\begin{figure}[hbt!]
  \begin{center}
    \mbox{
      \subfigure[]
{\scalebox{0.49}{
	  \includegraphics[angle=90,width=\linewidth]{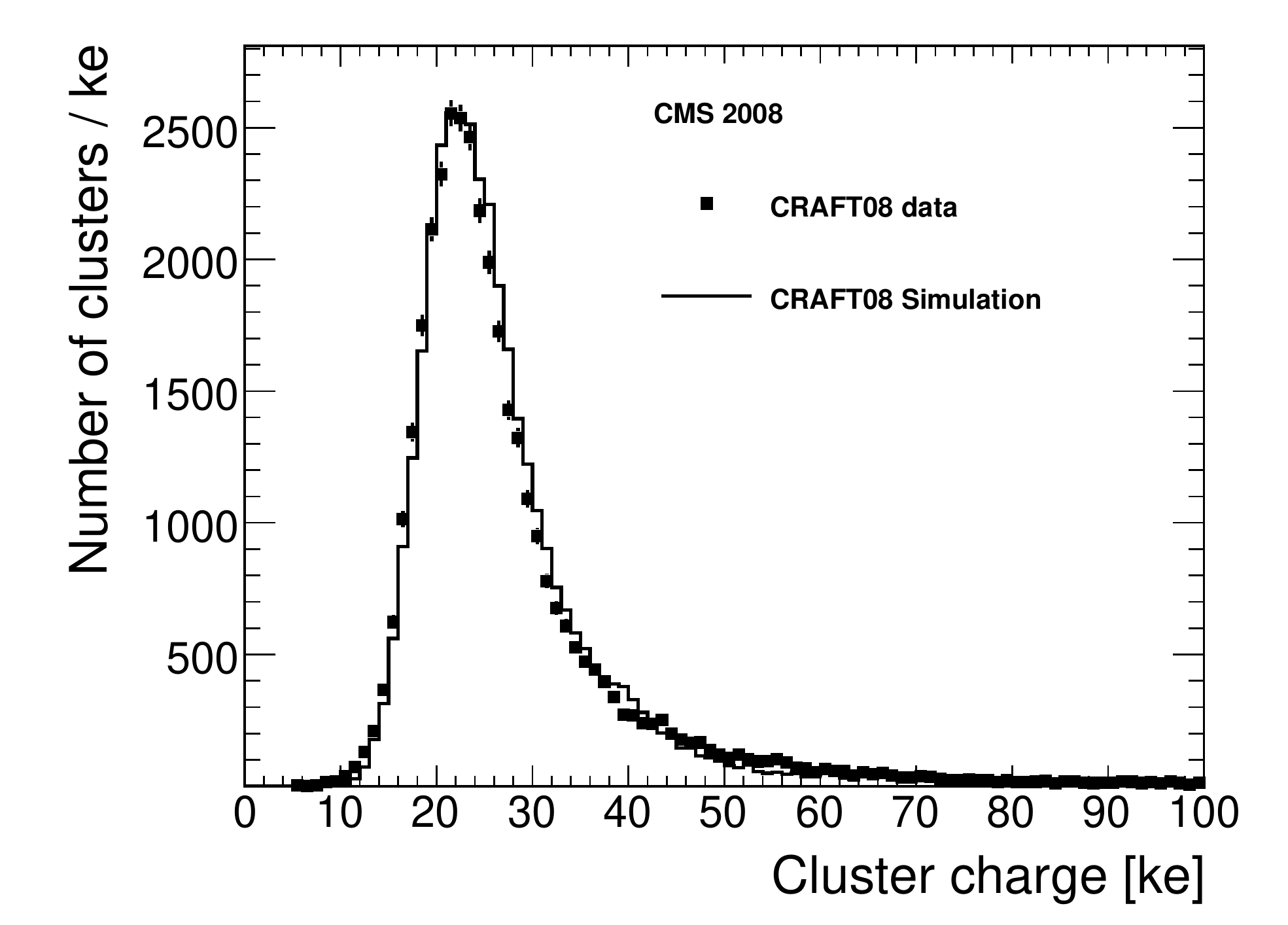}
	  \label{fig:clustercharge_BPIX} 
      }}
    }
    \hspace{-0.9cm}
    \mbox{
      \subfigure[]
{\scalebox{0.49}{
	  \includegraphics[angle=90,width=\linewidth]{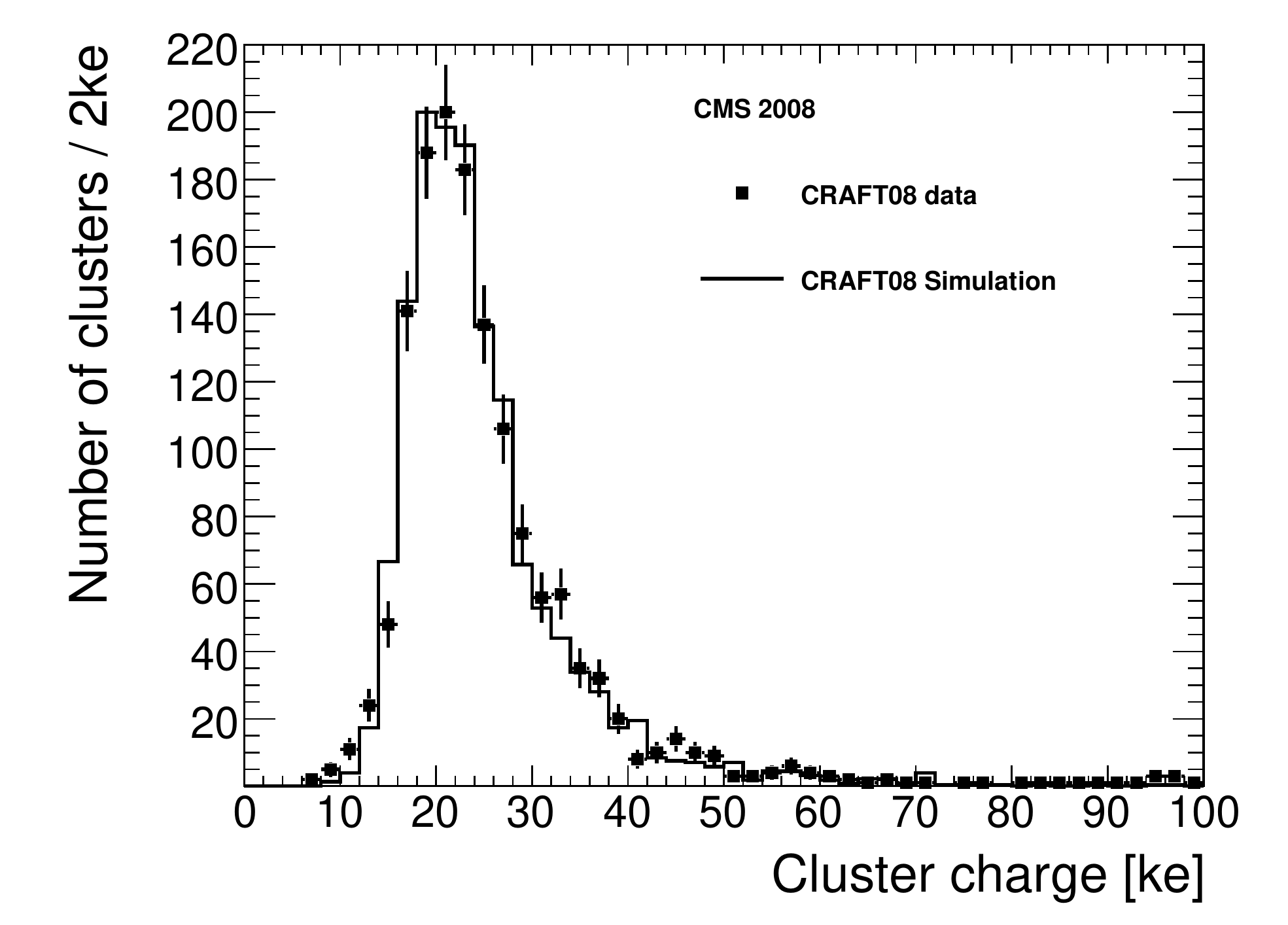}
	  \label{fig:cluster_charge_FPIX} 
      }}
    }

    \caption{Charge distribution in $10^3$ electrons (ke) for clusters larger than one pixel measured with the barrel (a) and endcap (b) pixel detector. The data points show the measurement with cosmic ray muons and the solid line the {\tt CMSCGEN} simulation. The simulated distributions are scaled to the data as described in the text.\label{fig:cluster_charge}}
  \end{center}
\end{figure}

Figure~\ref{fig:cluster_charge} shows the simulated and measured cluster charge after correcting for the track incidence angle. To emulate the angle distribution expected for collisions, tracks with transverse impact angle larger than 12$^\circ$ from the normal to the sensor surface are excluded from the study. 
Clusters are required to include at least two pixels and those with edge pixels are excluded from the sample. Finally, hits are excluded if more than one cluster is found within the same module or plaquette. The angular distribution of reconstructed tracks was compared with the {\tt CMSCGEN} predictions and found to be in good agreement~\cite{tracking}. 

To derive the most probable cluster charge and the width of the cluster charge distribution, fits to a Vavilov function~\cite{Vavilov:1957zz} are performed. The Vavilov function can be used to describe the energy deposition in thicker absorbers. The fit results are summarized in Table~\ref{tab:cluster_charge}. In addition to the cluster charge most probable value, the width, and the $\chi^2/\mbox{ndof}$ of the fit, the parameter $\kappa$ is stated in Table~\ref{tab:cluster_charge}. In the case that $\kappa$ is very small (e.g. $<0.01$) the Vavilov function converges to a Landau function, for large values of $\kappa$ (e.g. $>10$) the function converges to a Gaussian. The widths of the charge distribution agrees well with the simulation. The simulated charge peak is shifted by 1300 and 1000 electrons in the barrel and endcap respectively.  
The discrepancy is related to the uncertainty on the scale factors applied to the data
when converting the injected charge units into electrons.  The simulation predicts the charge from the deposited energy in the active region of the simulated sensor, using the ionization energy of silicon. In Fig.~\ref{fig:cluster_charge} the simulated charge has been shifted by the observed difference in peak positions.

\begin{table}[htb]
    \caption{Most probable value (MPV), width, $\kappa$ and $\chi^2/\mbox{ndof}$ of the Vavilov function fitted to the measured and simulated cluster charge distributions. Errors represent the uncertainties of the fit.}
    \begin{center}
        \begin{tabular}{lcccccccc} \hline
        	    & \multicolumn{4}{c}{Data ($10^3$ electrons)} & \multicolumn{4}{c}{Simulation ($10^3$ electrons)} \\
             & MPV & Width & $\kappa$ & $\frac{\chi^2}{\mbox{ndof}}$ & MPV & Width & $\kappa$ &  $\frac{\chi^2}{\mbox{ndof}}$ \\[4pt] \hline 
             Barrel & 23.9$\pm$0.2 & 3.7$\pm$0.1 & 0.18$\pm$0.02 & 1.6 & 22.6$\pm$0.2 & 3.4$\pm$0.1 & 0.13$\pm$0.02 & 1.6\\ 
              Endcap & 21.5$\pm$1.0 & 3.3$\pm$0.8 & 0.09$\pm$0.1 & 0.9 & 20.5$\pm$0.4 & 2.7$\pm$0.3 & 0.06$\pm$0.05 & 0.7\\ \hline
          \end{tabular}
      \end{center}
      \label{tab:cluster_charge}
  \end{table}

\subsection{Lorentz angle measurement}
In the presence of combined electric and magnetic fields, the drift of the charge carriers is affected by the Lorentz force. In the case of $n$-on-$n$ sensors electrons are collected at the $n^+$ pixel implant. The charges drift at an angle (Lorentz angle) relative to the direction of  the electric field, which leads to charge sharing among neighboring pixels. The pixel hit reconstruction exploits this effect to improve the spatial resolution by interpolating the charge collected in a cluster.  Once the interpolation is done the resulting position is adjusted to account for the Lorentz drift. Because the pixel barrel sensor planes are parallel to the magnetic field, the Lorentz drift is both maximal and in the azimuthal direction.  The forward pixel sensors are deliberately rotated by 20$^\circ$ with respect to the detector radial axes to produce a radial Lorentz drift and to increase azimuthal charge sharing.

\begin{figure}[htb]
  \begin{center}
    \includegraphics[width=9cm]{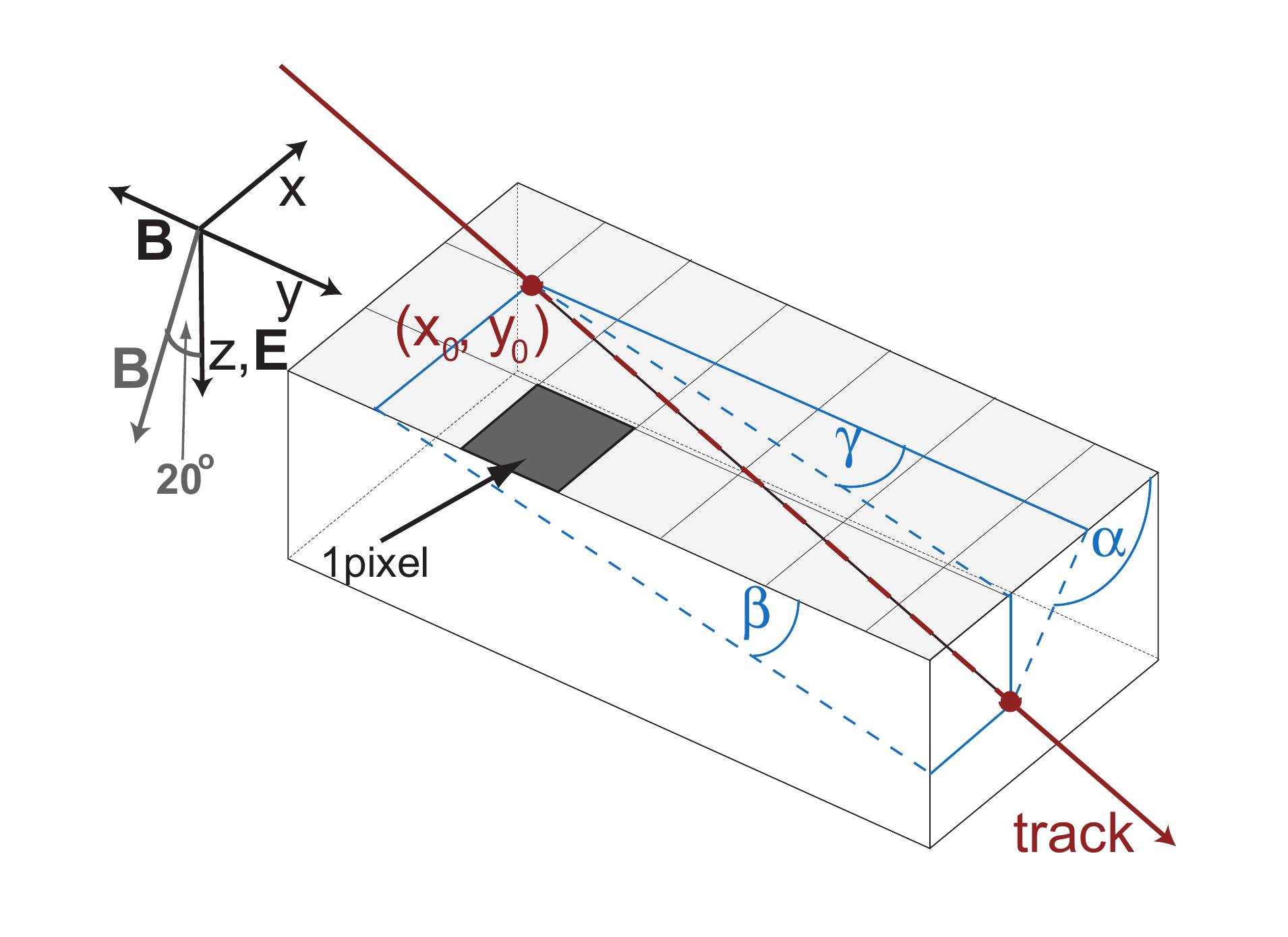}
    \caption{Sketch of the track impact angles with respect to a pixel sensor (pixel cell dimensions not to scale). The axes represent the local coordinate system. The magnetic field vector is anti-parallel to the $y$ axis for the barrel sensors and at 20$^\circ$ with respect to the $z$ axis for the endcap sensors.\label{fig:la_angles}}
  \end{center}
\end{figure}

The Lorentz angle extraction from the cosmic ray data is based on the cluster size method. The spread of the charge over neighboring pixels depends on the particle's incidence angle and has a minimum for tracks parallel to the drift direction of the charge carriers. The Lorentz angle is extracted by finding the minimum of the mean cluster size along the local $x$ coordinate measured as a function of the cotangent of the incidence angle $\alpha$, as shown in Fig.~\ref{fig:la_angles}. The fit function is given by
\begin{equation}
f(\xi) = p_0 + \sqrt{ p_1^2 + p_2^2 (\xi - \xi_{\textrm {min}})^2 }\quad ,
\end{equation}
where $\xi=\cot(\alpha)$, the parameter $\xi_{\textrm{ min}}$ is the location of the function minimum, and $p_2$ is independently fitted for $\xi$ values larger and smaller than the function minimum.

In this analysis the track $\chi^2$ per degree of freedom, $\chi^2/{\textrm {ndf}}$, is required to be smaller than 2, and clusters with double-sized or edge pixels are excluded. In addition, to reject residual single-pixel hits from out-of-time particles, only clusters with at least two pixels along the local $y$ coordinate are accepted. 

\begin{figure}[hbt]
  \begin{center}
    \mbox{
      \subfigure[]
{\scalebox{0.41}{
	  \includegraphics[width=\linewidth]{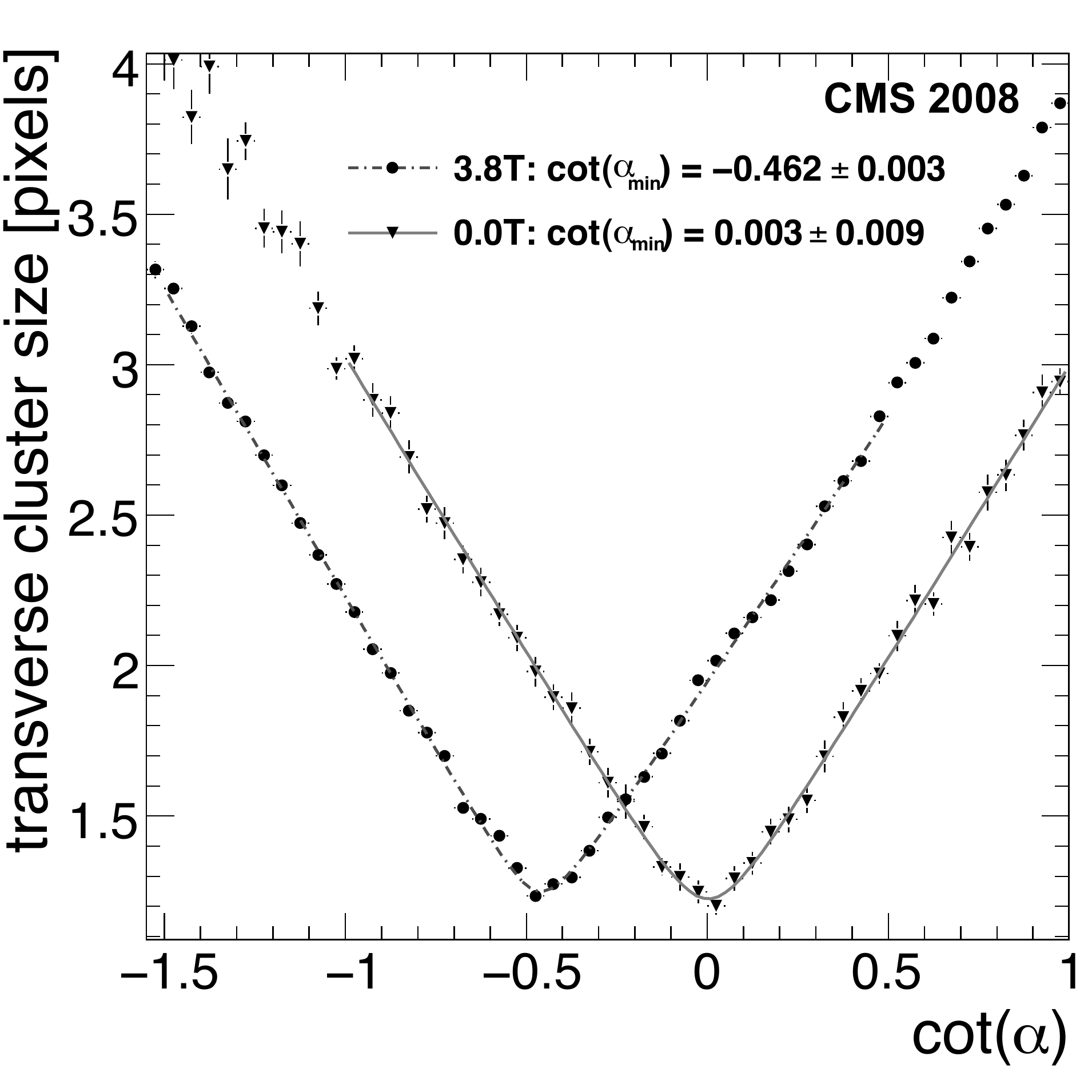}
	  \label{fig:LA_BPIX} 
      }}
    }
    \mbox{
      \subfigure[]
{\scalebox{0.41}{
	  \includegraphics[width=\linewidth]{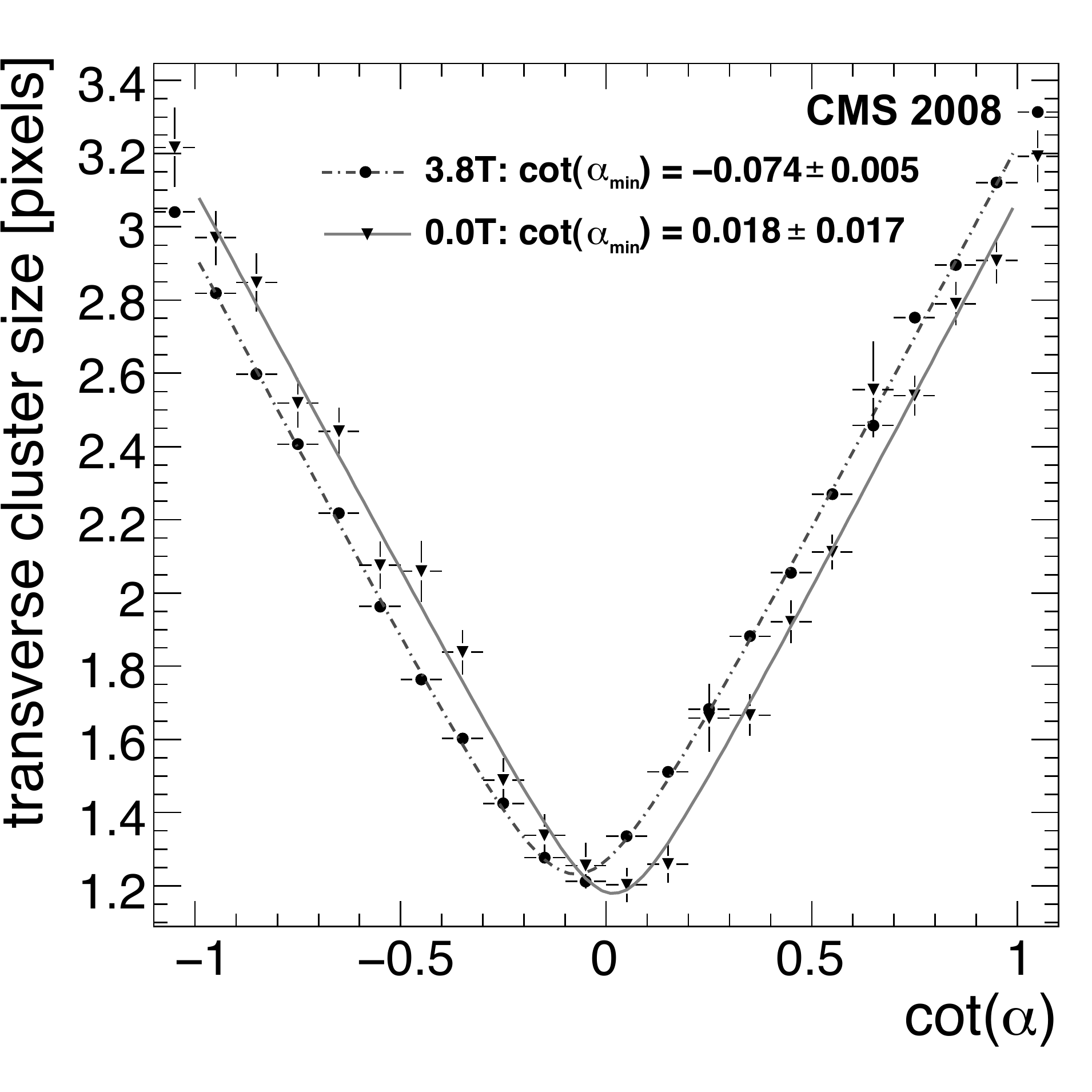}
	  \label{fig:LA_FPIX} 
      }}
      \label{fig:LA}
    }
    \caption{Cluster size along the local $x$ axis as function of the cotangent of the impact angle $\alpha$ measured in the barrel (a) and endcap (b) regions of the pixel detector. The circles correspond to the measurement with the 3.8~T field while the triangles correspond to the measurement without magnetic field. The dashed line shows the fit to the data points.}
  \end{center}
\end{figure}

Figure~\ref{fig:LA_BPIX} shows the transverse cluster size as a function of $\cot(\alpha)$  in the barrel pixel detector with 3.8~T and without field. The minimum $\xi_{\textrm {min}}$ obtained for the data at 3.8~T is $-0.462\pm0.003$. 
The measured value corresponds to a Lorentz angle of (24.8$\pm$0.2)$^\circ$ and a shift of the transverse hit position of 65.9$\pm$0.5 $\mu$m for a 285 $\mu$m thick barrel sensor. The quoted uncertainty is purely statistical.
The systematic uncertainties on $\xi_{\textrm {min}}$ related to the selection cuts and fit range are estimated to be approximately 3\%. 


The measured Lorentz angle value may be compared with the prediction of the {\tt PIXELAV} program~\cite{Chiochia:2004qh,Swartz:2005vp}. In the program the value of the reverse bias voltage is set to 100~V and 300~V for the barrel and endcap detectors, respectively. The sensor temperature and the Hall factor are set to 20~$^\circ$C and 1.02, respectively. The dependence of the charge carrier mobility on the electric field is taken from Ref.~\cite{jacoboni}. The systematic errors on the predicted values  are dominated by the uncertainty on the Hall mobility and can be as large as 10\%. The prediction for the barrel of $\xi = -0.452\pm0.002$ agrees well with the measured value.

To check the correctness of the method, the measurement is performed also using data collected without magnetic field. The corresponding minimum is found to be 0.003$\pm$0.009, consistent with zero, as expected in the absence of magnetic field.

Figure~\ref{fig:LA_FPIX} shows the transverse cluster size as function of $\cot(\alpha)$ for the endcap section. The minimum obtained with the 3.8~T field is $-0.074\pm0.005$ and also agrees well with the {\tt PIXELAV} prediction of $-0.074\pm0.004$. The measured value corresponds to a Lorentz angle of (4.2$\pm$0.3)$^\circ$ and a hit position shift of 9.9$\pm$0.7 $\mu$m for the 270 $\mu$m thick endcap sensors.
The minimum obtained with no magnetic field is consistent with zero (0.018$\pm$0.017). 

The measured values of the Lorentz angle were used when reprocessing the collected and simulated data samples and therefore the Lorentz effect was correctly taken into account while performing the spatial alignment of the pixel detector~\cite{alignment}.
\subsection{Hit detection efficiency\label{sec:hitefficiency}}
\begin{figure}[hbt]
  \begin{center}
    \mbox{
      \subfigure[]
{\scalebox{0.325}{
	  \includegraphics[width=\linewidth]{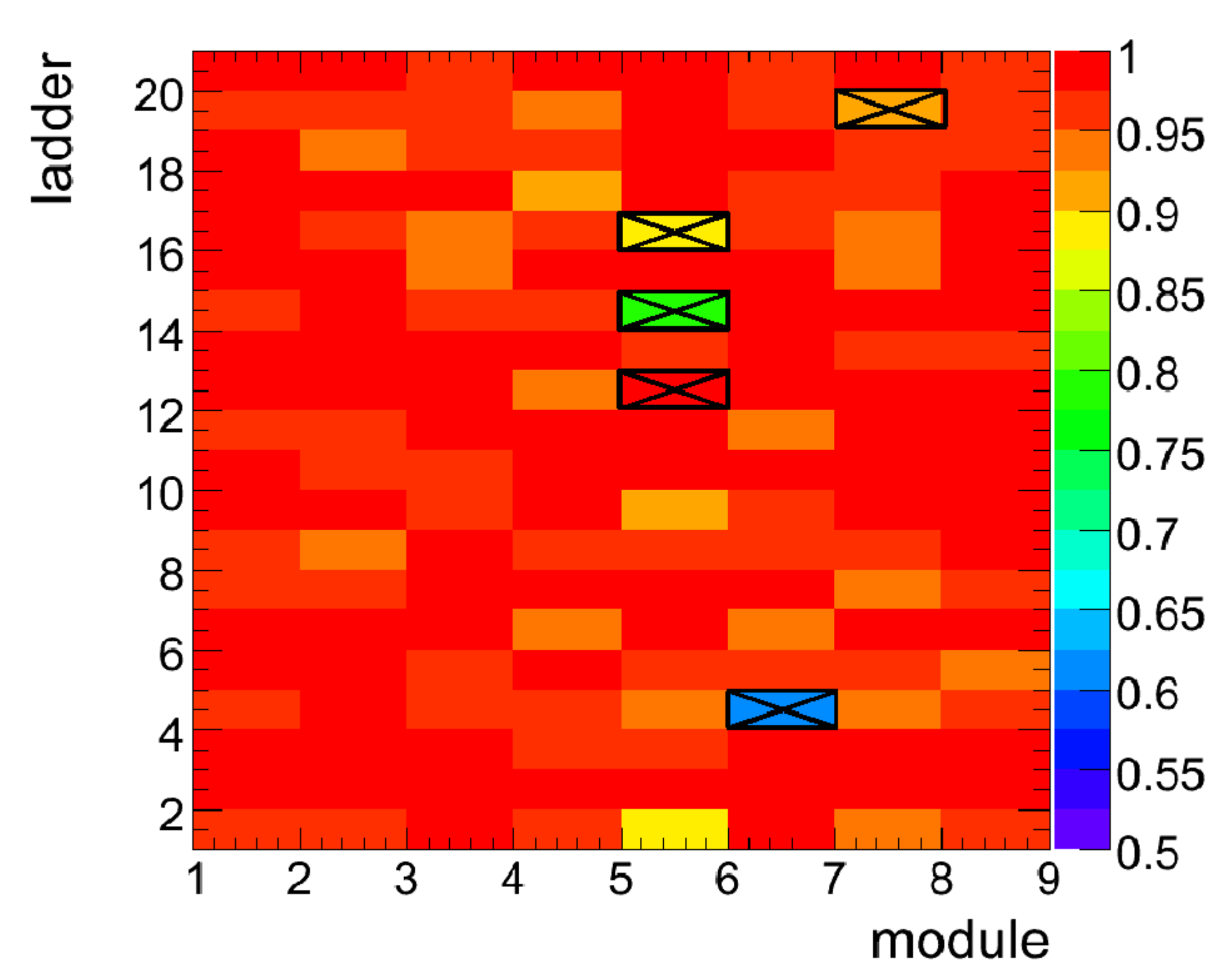}
	  \label{fig:eff_lay1} 
      }}
    }
    \hspace{-4mm}
    \mbox{
      \subfigure[]
{\scalebox{0.325}{
	  \includegraphics[width=\linewidth]{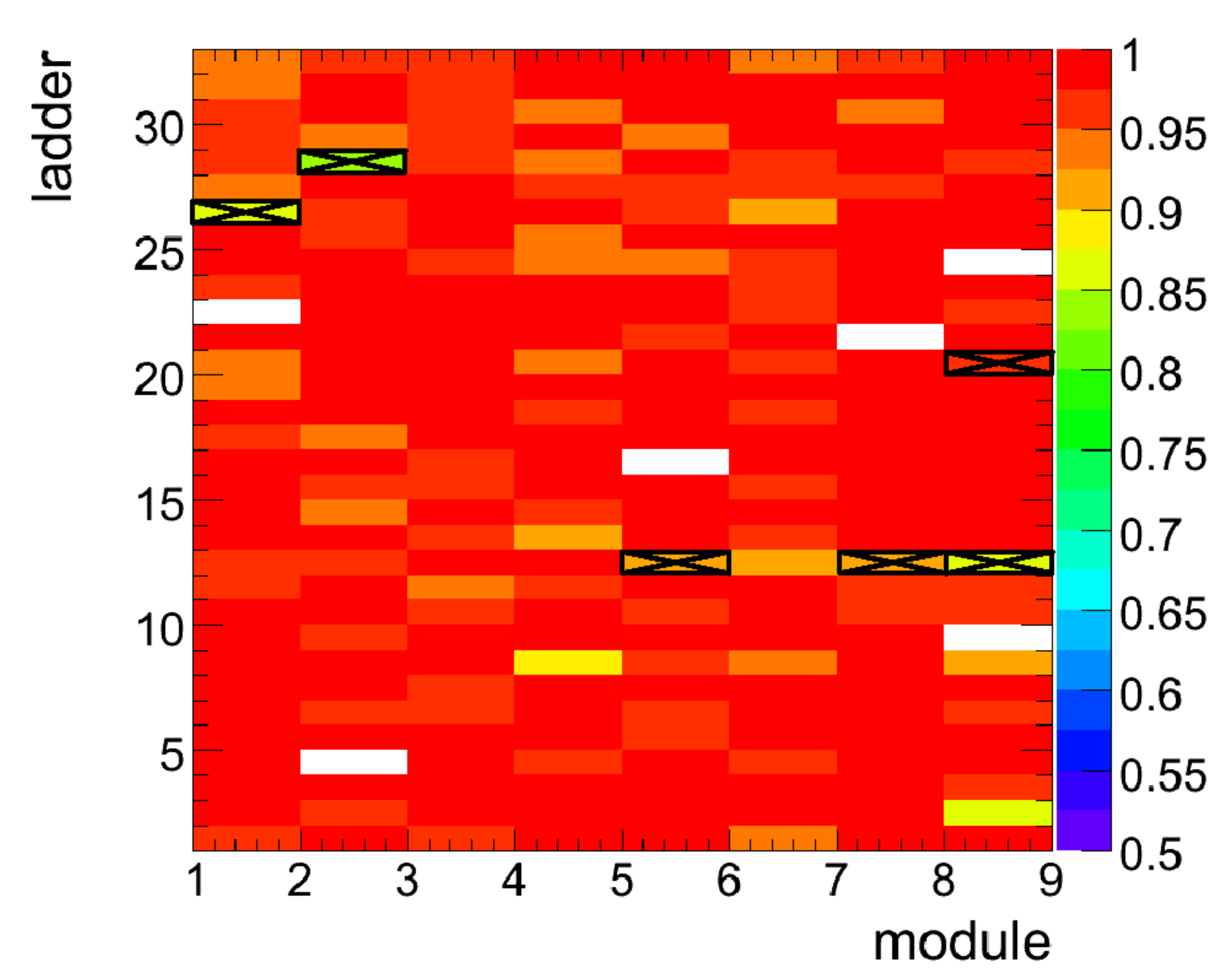}
	  \label{fig:eff_lay2} 
      }}
    }
    \hspace{-4mm}
    \mbox{
      \subfigure[]
{\scalebox{0.325}{
	  \includegraphics[width=\linewidth]{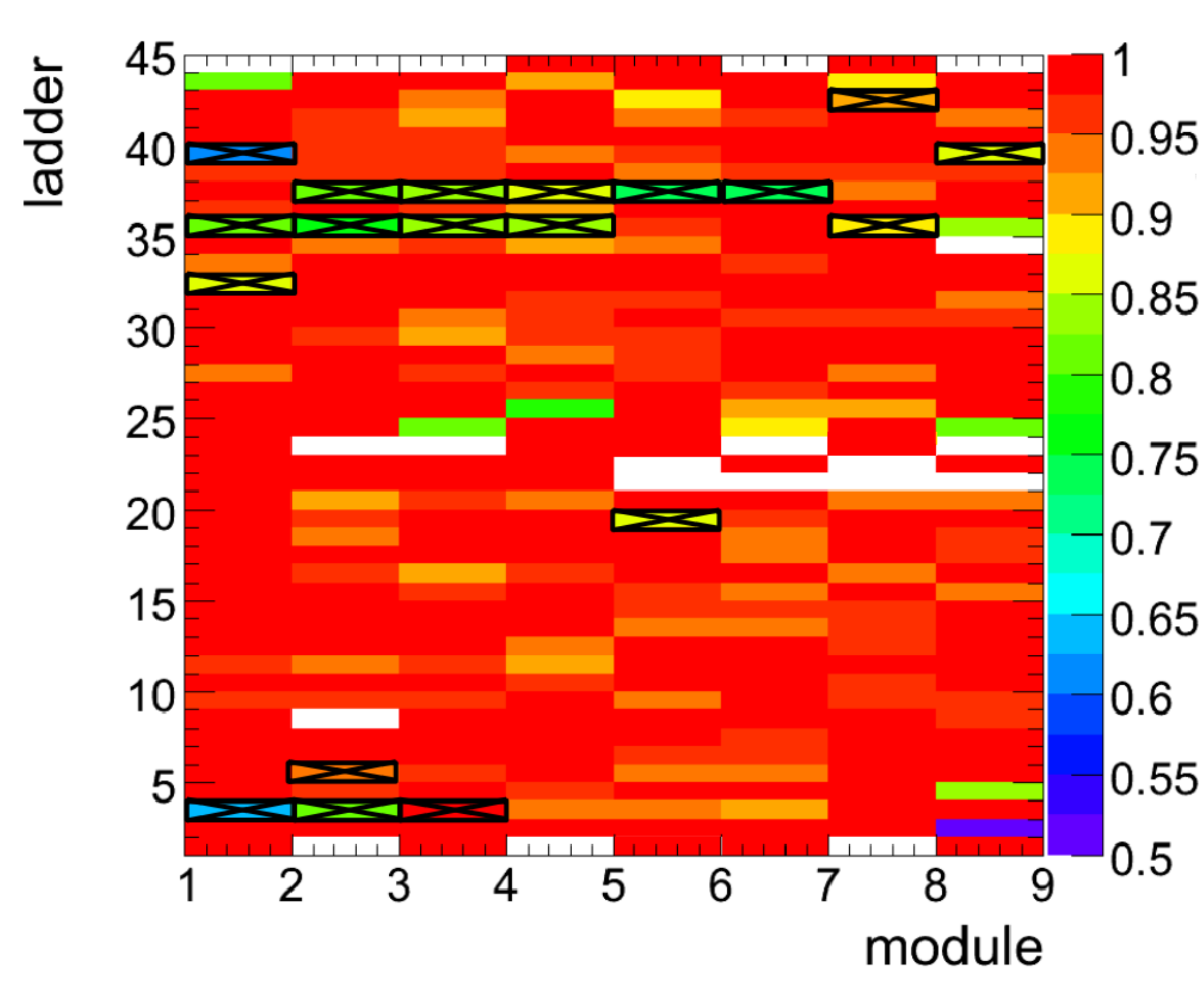}
	  \label{fig:eff_lay3} 
      }}
    }
\caption{Hit detection efficiency measured in each sensor of the first (a), second (b), and third (c) barrel layers. Modules affected by configuration problems, missing sensor bias or inactive ROCs are marked with a crossed black box.\label{fig:hitefficiency}}
  \end{center}
\end{figure}

The hit reconstruction efficiency is measured by extrapolating tracks to the pixel sensors and checking the presence of a compatible pixel hit~\cite{tracking,:2009mq}. If one is found, it is added to the track and the trajectory is updated with the new information. In this case the hit is called {\it valid}. If no hit is found in the search window, a {\it missing} hit is added to the trajectory. The hit reconstruction efficiency is defined as
\begin{equation}
\epsilon = \frac{\sum valid}{\sum (valid+missing)}\quad.
\end{equation}
The following selection cuts are applied:
\begin{itemize}
\item For each measurement one additional valid hit was required to be associated to the track in both top and bottom halves of the pixel detector with respect to the beam plane;
\item The muon arrival time at the pixel detector is required to be
  within 5 ns of the time that has maximum efficiency, to take into
  account the timing alignment between the pixel and muon systems;
\item An extrapolated hit, either valid or missing, is not counted in the efficiency calculation when its distance to the sensor edge is smaller than the uncertainty on the track trajectory propagated to the sensor surface; 
\item Only events with a single track with momentum larger than 10~GeV/$c$ are accepted.
\end{itemize}
The position of the muon timing cut window is centered on the maximum
efficiency, and adjusted by the 9 ns timing shift between the two subsets
of CRAFT data (Sec.~\ref{sec:dataset}).  This offset turns out to be
-8 ns and +1 ns for the early and late subsets, respectively.  
As the cut on the muon timing window is
relaxed, the efficiency decreases, consistent with tracks arriving
outside the time acceptance of the pixel readout chip, taking into
account the resolution of the muon time measurement.
The efficiency measured in each module of the three barrel layers is shown in Fig.~\ref{fig:hitefficiency}. The number of selected hits in the endcap region is not sufficient to perform this study. 
Modules with efficiency below 90\% correlate with known configuration problems, missing sensor bias, or inactive ROCs. These modules are represented by cells marked by a crossed black box in Fig.~\ref{fig:hitefficiency}. Modules excluded from the DAQ or with insufficient number of hits are represented by white cells. Removing these modules from the measurement, the layer efficiency averaged over the modules is (97.1$\pm$1.4)\%, (97.1$\pm$1.9)\% and (96.4$\pm$2.6)\% in the first, second, and third barrel layers, respectively, where the statistical error is the RMS spread. 

The measured hit detection efficiencies after the cuts are lower than test beam measurements~\cite{Allkofer:2007ek}. A correlation between the efficiency and the track impact angle is observed, showing lower efficiency for tracks parallel to the charge carriers drift direction. These tracks are more likely to create single-pixel hits.
A loss of single-pixel hits in cosmic ray events can be explained by 
random arrival times of cosmic rays with respect to the beam crossing clock
and the time-walk effect giving different time-stamps to high and low charge hits from the same track~\cite{Kastli:2005jj}. 
These effects are peculiar to data taking with cosmic ray particles and are not expected to affect proton collisions, where outgoing particles are synchronized with the machine clock. 

\subsection{Hit residuals and position resolution}

The silicon strip 
and pixel tracker are spatially aligned as described in Ref.~\cite{alignment}. After alignment, the hit residuals are measured by refitting each track and comparing for each hit the track prediction and the reconstructed hit position. The measured hit is excluded from the track fit, and only tracks with momenta larger than 10~GeV/$c$ are taken. 
Clusters of a single pixel are excluded if the cluster charge is below 10\,000 electrons, to avoid time-walk effects, which can split clusters from cosmic ray tracks between two beam crossings. 
In addition, clusters are excluded if they are close to the sensor edges, as described in Section~\ref{sec:hitefficiency}. The residual distributions are measured as a function of the impact angles and cluster charge and a Gaussian fit is performed in each bin. Contributions to the width of the residual distribution come from track extrapolation error, intrinsic detector resolution, and multiple scattering. The track extrapolation error is the largest component and includes the uncertainties due to residual misalignment of the sensors used in the trajectory extrapolation with respect to the measured sensor.

\begin{figure}[hbt]
	\begin{center}
    \mbox{
      \subfigure[]
{\scalebox{0.32}{
	  \includegraphics[angle=90,width=\linewidth]{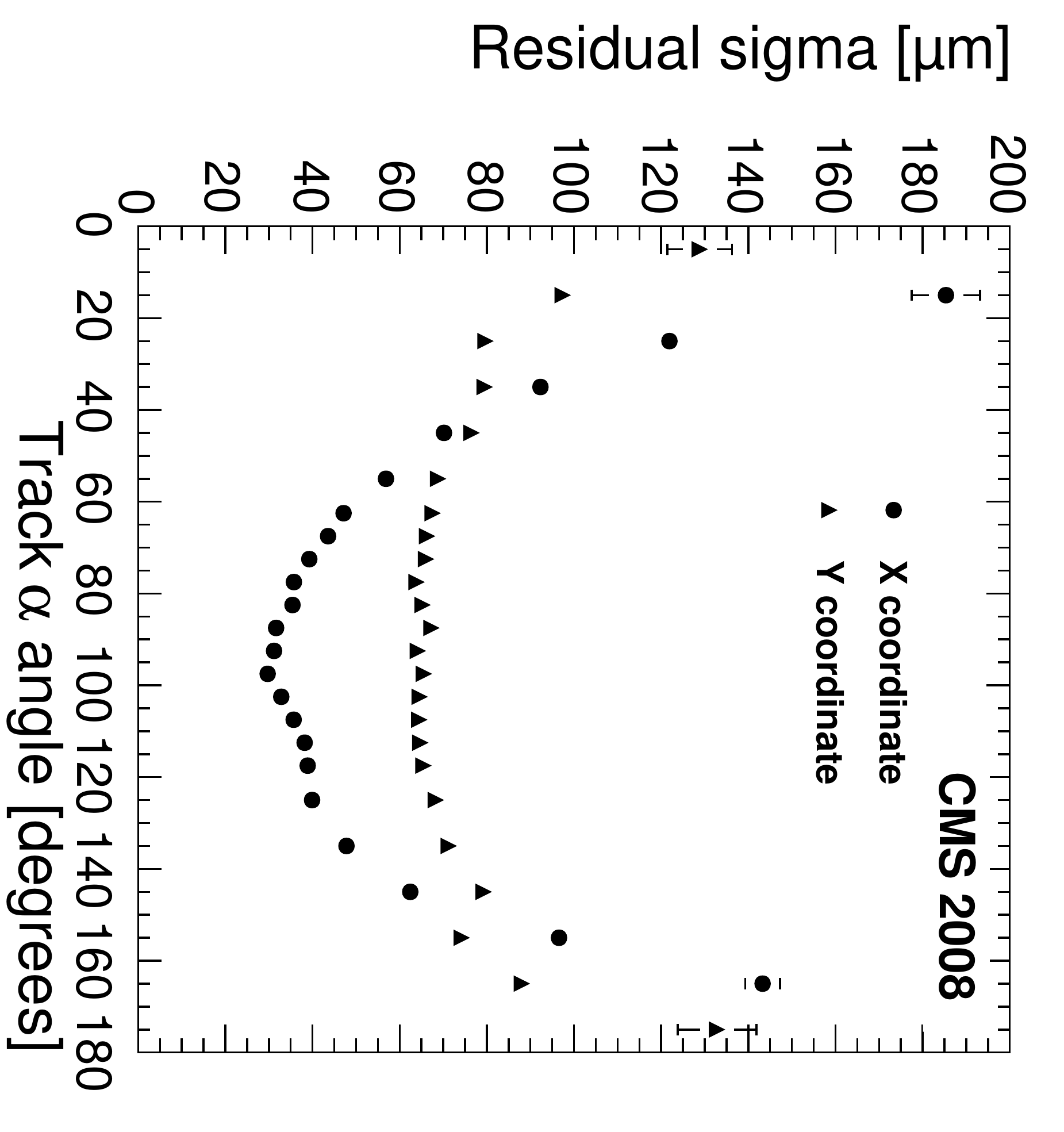}
	  \label{fig:residualsalpha} 
      }}
    }
    \hspace{-5mm}
    \mbox{
      \subfigure[]
{\scalebox{0.32}{
	  \includegraphics[angle=90,width=\linewidth]{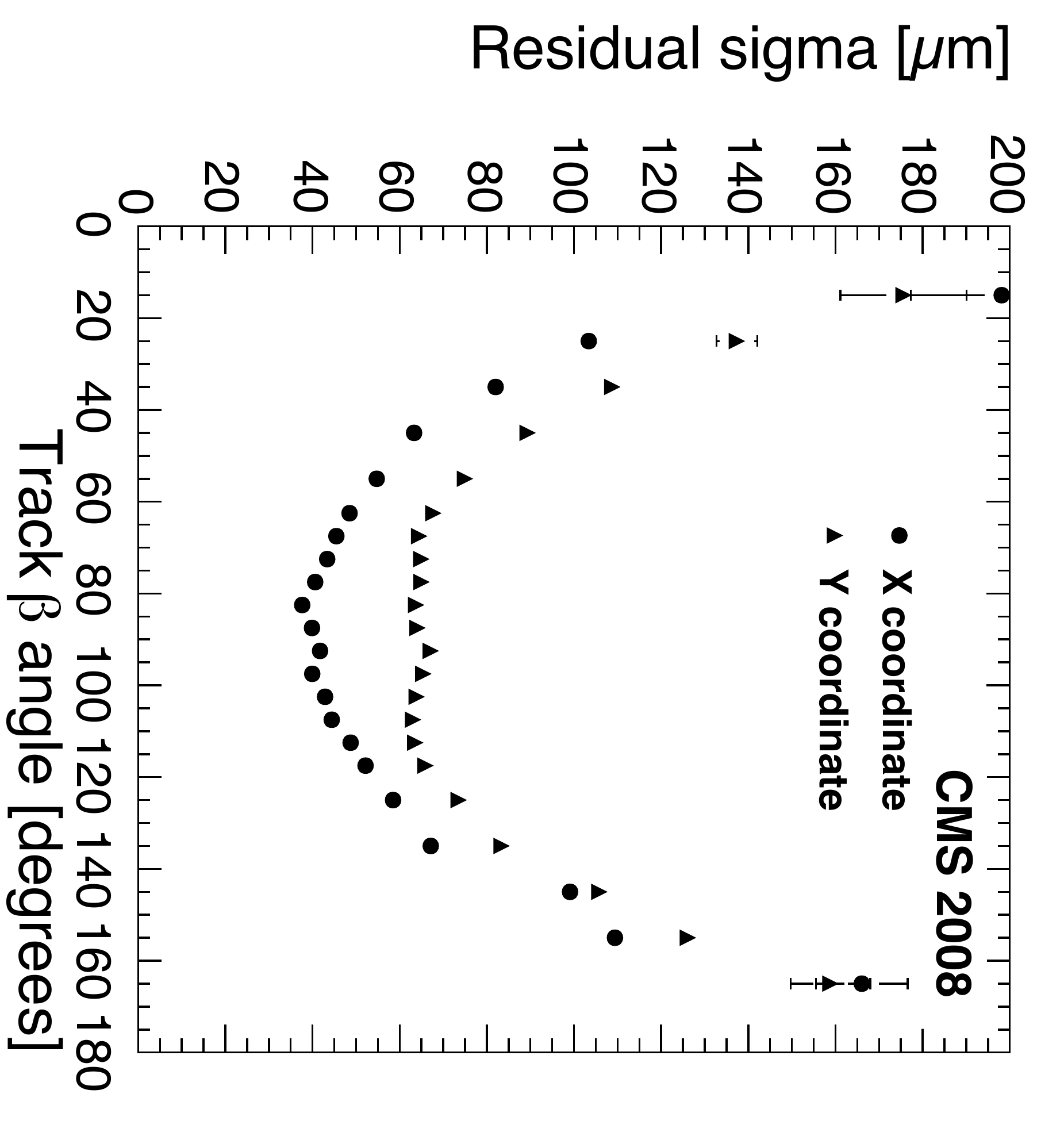}
	  \label{fig:residualsbeta} 
      }}
    }
    \hspace{-5mm}
    \mbox{
      \subfigure[]
{\scalebox{0.32}{
	  \includegraphics[angle=90,width=\linewidth]{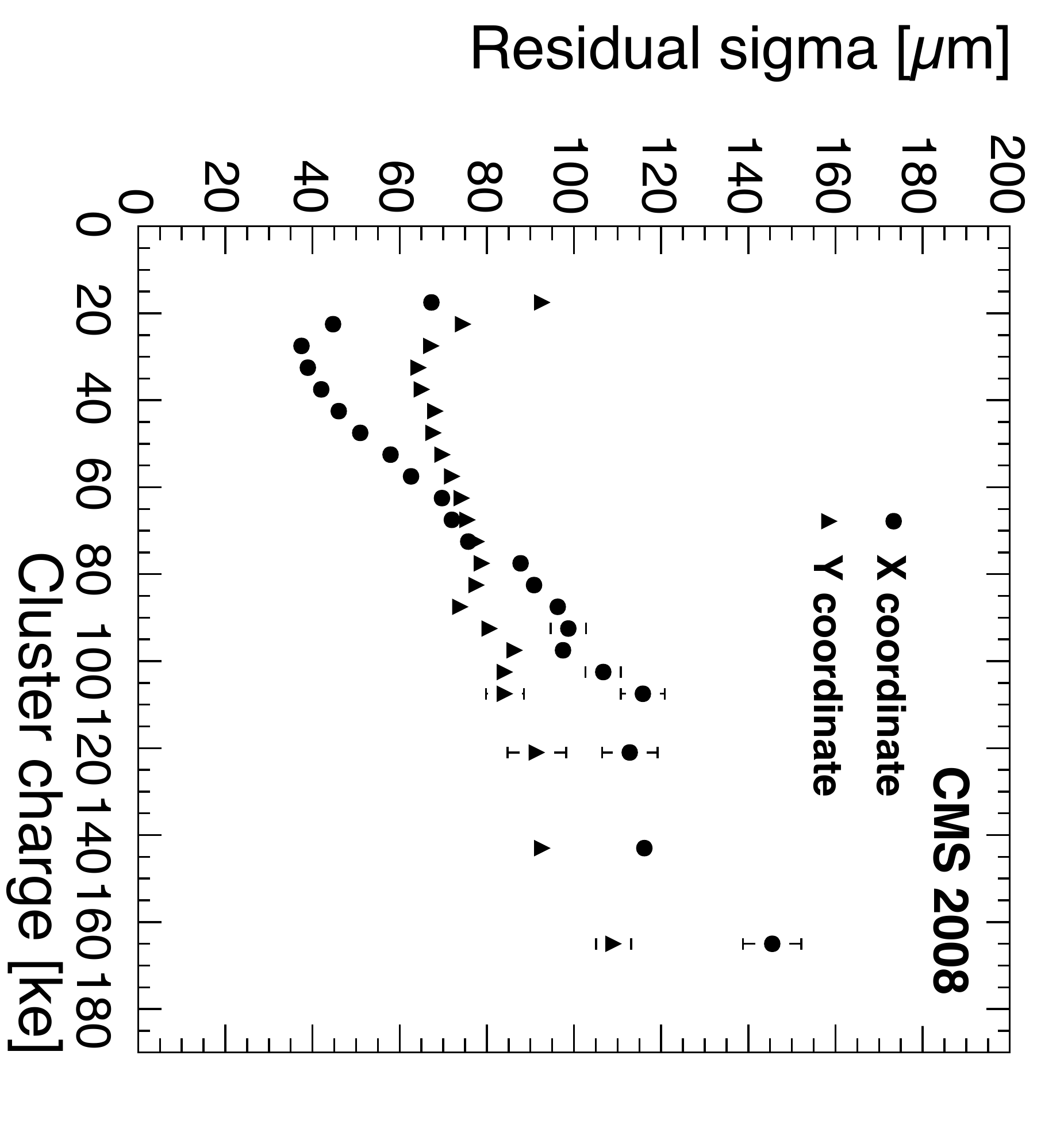}
	  \label{fig:residualscharge} 
      }}
    }
    \caption{Pixel barrel hit residuals as a function of the transverse impact angle $\alpha$ (a), longitudinal impact angle $\beta$ (b), and cluster charge in $10^3$ electrons (c). Circles and triangles correspond to the transverse ($x$) and longitudinal ($y$) local coordinates, respectively.
Normally incident tracks have impact angles of 90$^\circ$.
\label{fig:posresiduals}}
  \end{center}
\end{figure}

Figure~\ref{fig:posresiduals} shows the width from the Gaussian fit of the barrel residual distribution as a function of the $\alpha$ and $\beta$ angles (see Fig.~\ref{fig:la_angles}) and the total cluster charge. For tracks normal to the sensor plane, the Gaussian sigma is about 30 $\mu$m and 65 $\mu$m along the local $x$ and $y$ coordinates, respectively. The width strongly depends on the cluster charge and impact angles. The best precision is expected for clusters that have two-pixels-wide projections and charges of about 30\,000 electrons. For larger values of the cluster charge, the position resolution deteriorates due to delta rays. 

The detector intrinsic position resolution is measured using tracks that traverse overlapping sensors in the barrel layers. A detailed explanation of the measurement technique is given in Ref.~\cite{:2009mq}. Tracks passing through two overlapping modules in the same layer are used to compare the hit position with the expected position from the track trajectory. 
The difference of the local track impact points is about ten times more precise than the individual predicted hit positions and always below 5 $\mu$m. The {\it double difference} is formed by taking the difference between the hit position difference and the predicted position difference ($\Delta x_{\rm{pred}}$). The width of this double difference distribution is insensitive to translational misalignment of the overlapping modules.

\begin{figure}[hbt]
  \begin{center}
    \mbox{
      \subfigure[]
{\scalebox{0.47}{
\includegraphics[angle=90,width=\linewidth]{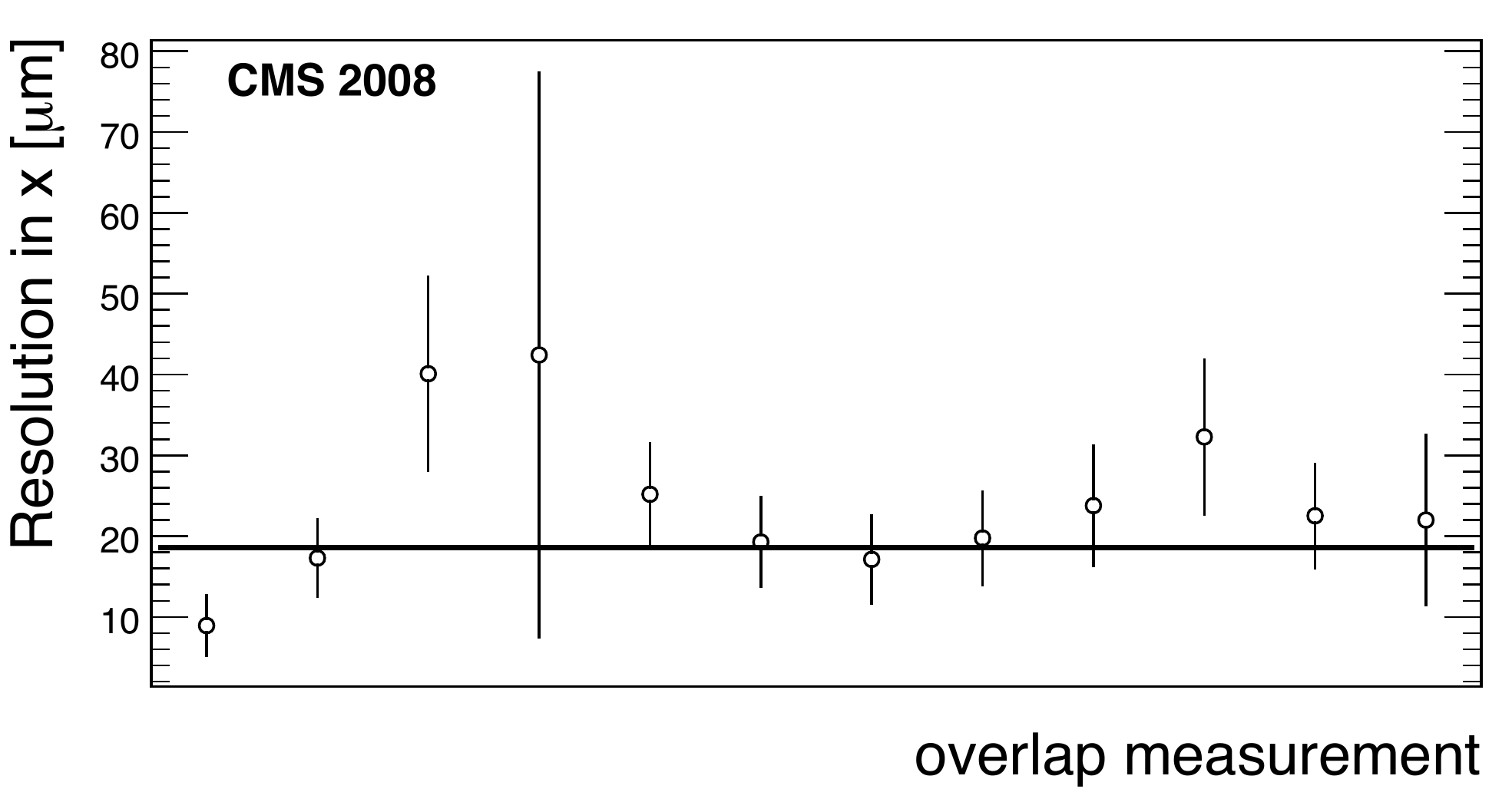}
	  \label{fig:intrinsic_res_localx} 
      }}
    }
    \mbox{
      \subfigure[]
{\scalebox{0.47}{
\includegraphics[angle=90,width=\linewidth]{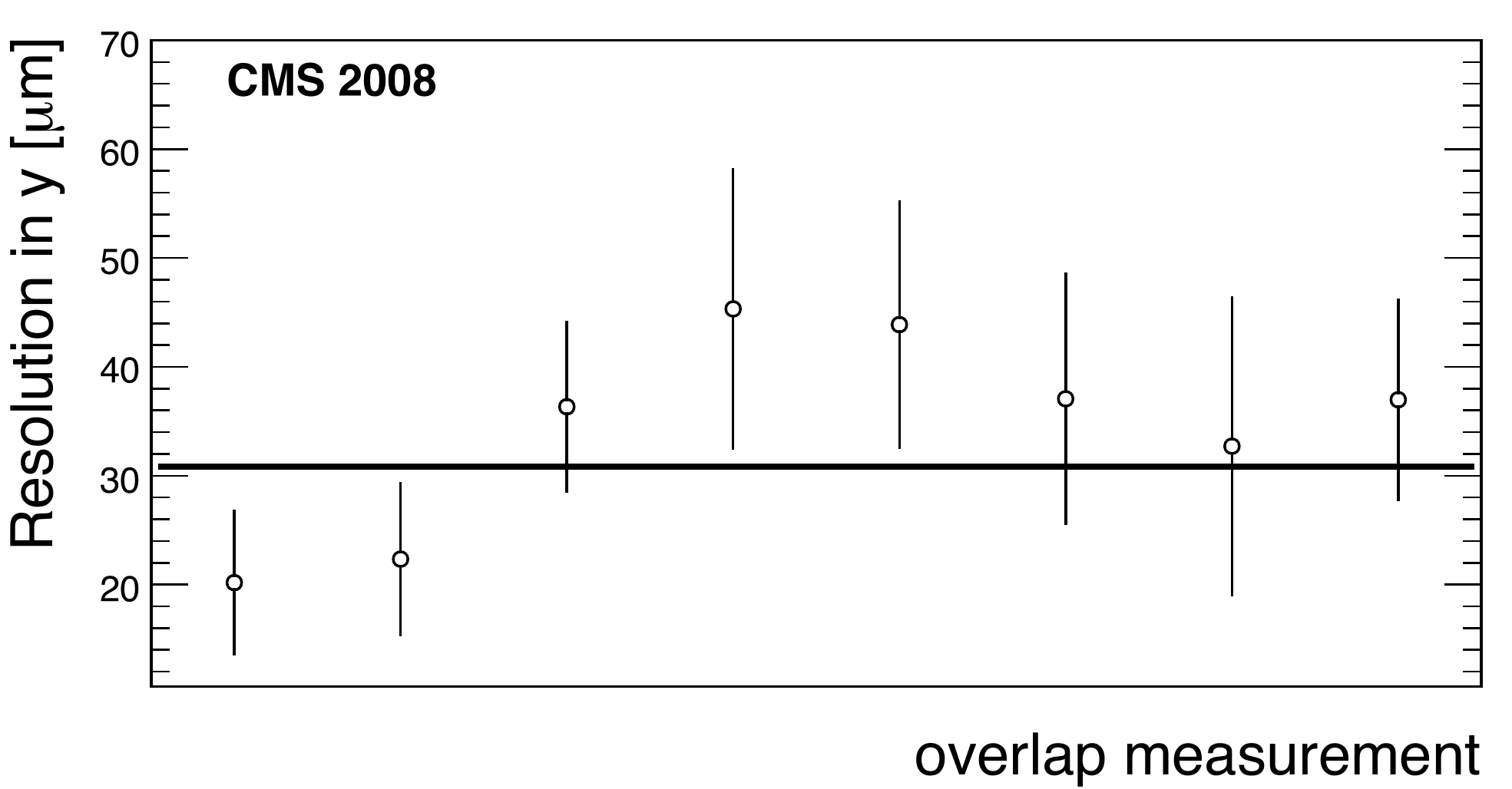}
	  \label{fig:intrinsic_res_localy} 
      }}
    }

    \caption{Hit position resolution of the barrel pixel detector along the local transverse (a) and longitudinal (b) coordinates measured using overlapping sensors within a layer. Each measurement point corresponds to a different pair of overlapping modules in $z$ for (a) and in $r$-$\phi$ for (b). The circles show the measurement while the solid line represents the error-weighted mean of the measurements.\label{fig:intrinsic_res}}
  \end{center}
\end{figure}

The study is performed for both directions, $x$  and $y$, in the pixel
module coordinates. To limit the effect of multiple scattering, a
minimum momentum cut of 5~GeV/$c$ is applied.  Clusters with charge
below 10\,000 electrons or containing pixels on the sensor edge are 
excluded.  The charge cut removes only 2.6\% of all clusters.
Tracks with angles greater than 30$^\circ$ from the normal
are also excluded. The double difference widths are fitted with a
Gaussian and the uncertainty from the tracking prediction is
subtracted quadratically to recover the hit resolution on the position
difference. The results for the resolution are shown in
Fig.~\ref{fig:intrinsic_res} for the local $x$ and $y$
coordinates. Each data point represents a measurement extracted from a
pair of overlapping sensors with at least 30 crossing tracks. 

The solid lines are weighted means of the measured resolutions. With the assumption of equal resolution for each of the modules in the overlap, the final fit values for the resolution for a single module are
$18.6\pm1.9$ $\mu$m along $x$ and $30.8\pm3.2$ $\mu$m along $y$.
Events with the same range of impact angles as in the measured sample are simulated using the {\tt PIXELAV} program. This simulation does not include misalignment effects. The residual distributions are obtained comparing the true and reconstructed hit positions. A Gaussian fit to the simulated residual distribution gives a resolution of 22.1$\pm$0.2 $\mu$m and 28.5$\pm$0.1 $\mu$m along the local $x$ and $y$ coordinates, respectively, in good agreement with the measurements.

The position resolution will be further improved by reducing and equalizing the readout thresholds before operation with colliding beams.
In test beam measurements, an RMS resolution of 12~$\mu$m was measured along the local coordinate $x$ affected by the Lorentz drift, using tracks perpendicular to the sensor, a readout threshold of 2500 electrons, and a 3~T magnetic field~\cite{Alagoz:2009}.

\subsection{Track parameter residuals at vertex~\label{sec:trackparresolution}}
The resolution on the track impact parameters is an important benchmark of the pixel system. After detector alignment, the track parameter resolution can be extracted from data as described in Ref.~\cite{alignment}.  The method splits the cosmic ray tracks in the barrel section at the point of closest approach to the nominal beamline, creating two track candidates (or {\it legs}). The top and bottom legs are taken as two independent tracks and fitted accordingly, with the track parameters propagated to their respective points of closest approach to the beamline.  The transverse residual distribution is defined as: $\Delta d_{xy} = (d_{xy,1} - d_{xy,2})/\sqrt{2}$, where the factor of $\sqrt{2}$ is included due to statistically independent legs and $d_{xy,1}$, $d_{xy,2}$ are the track transverse impact parameters of each track leg. The residual of the longitudinal impact parameter, $d_z$, is defined similarly.
The measurement technique was validated using a full detector simulation.
To select a sample that closely resembles tracks expected from collision events, each track leg is required to have a momentum larger than 4~GeV/$c$, at least 10 hits in the tracker, of which at least two two-dimensional hits are in the strip tracker, and three hits are in the barrel pixel section. In addition, the track $\chi^2$ is required to be smaller than 100. 
\begin{figure}[hbt]
  \begin{center}
    \mbox{
      \subfigure[]
{\scalebox{0.47}{
	  \includegraphics[width=\linewidth]{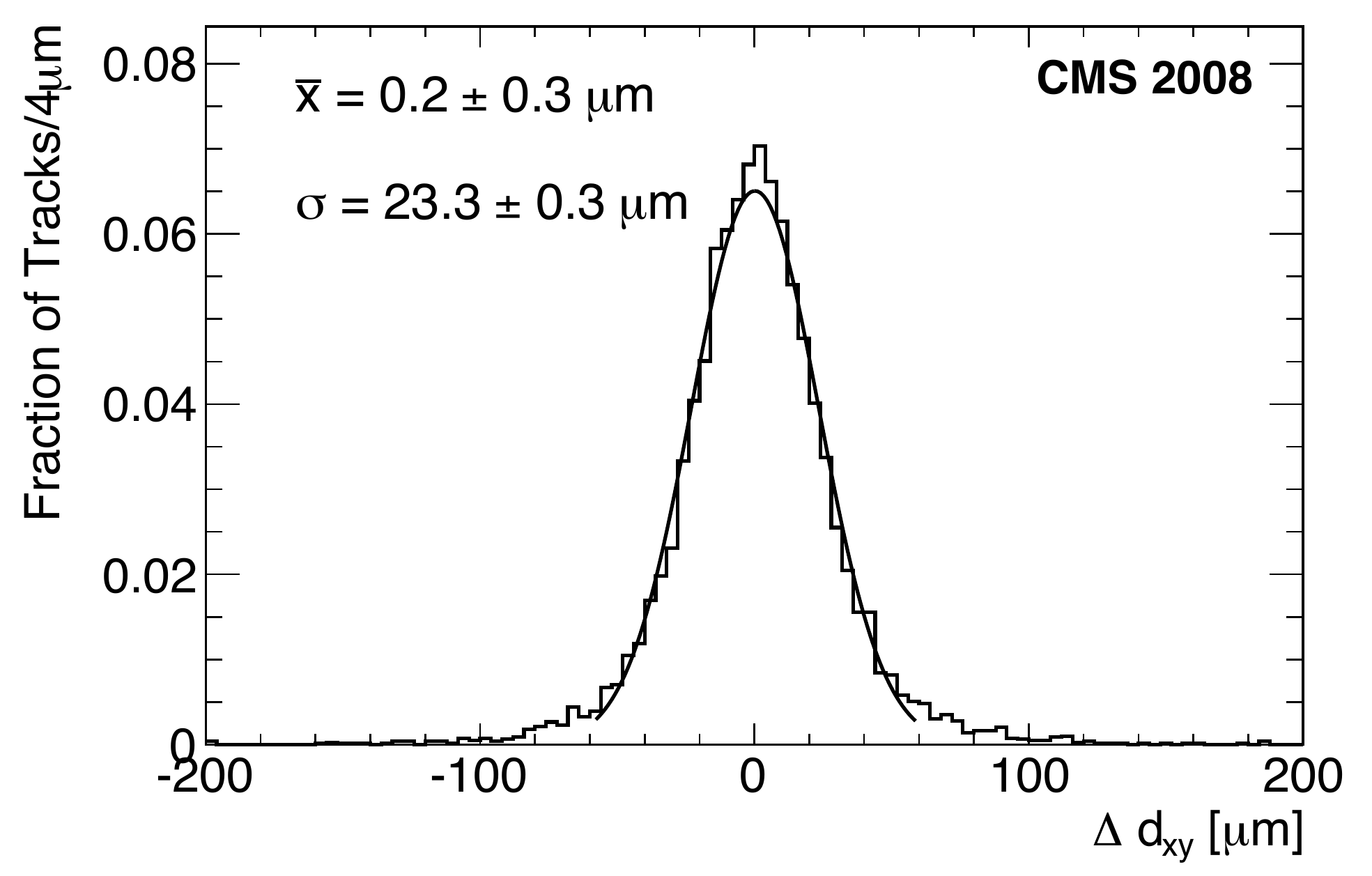}
	  \label{fig:dxy_residuals} 
      }}
    }
    \hspace{-0.5cm}
    \mbox{
      \subfigure[]
{\scalebox{0.47}{
	  \includegraphics[width=\linewidth]{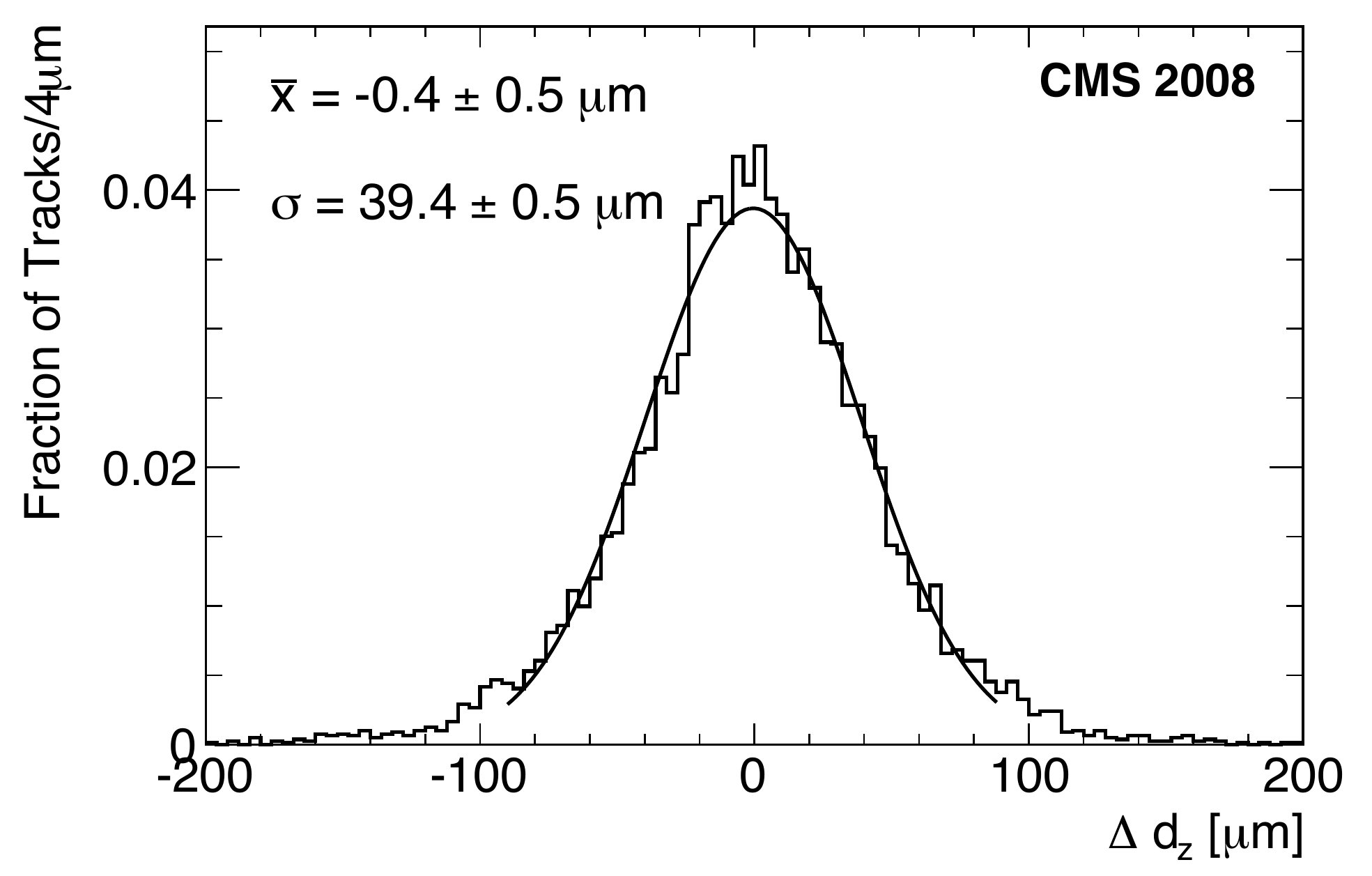}
	  \label{fig:dz_residuals} 
      }}
      \label{fig:impact_res}
    }

    \mbox{
      \subfigure[]
{\scalebox{0.47}{
	  \includegraphics[width=\linewidth]{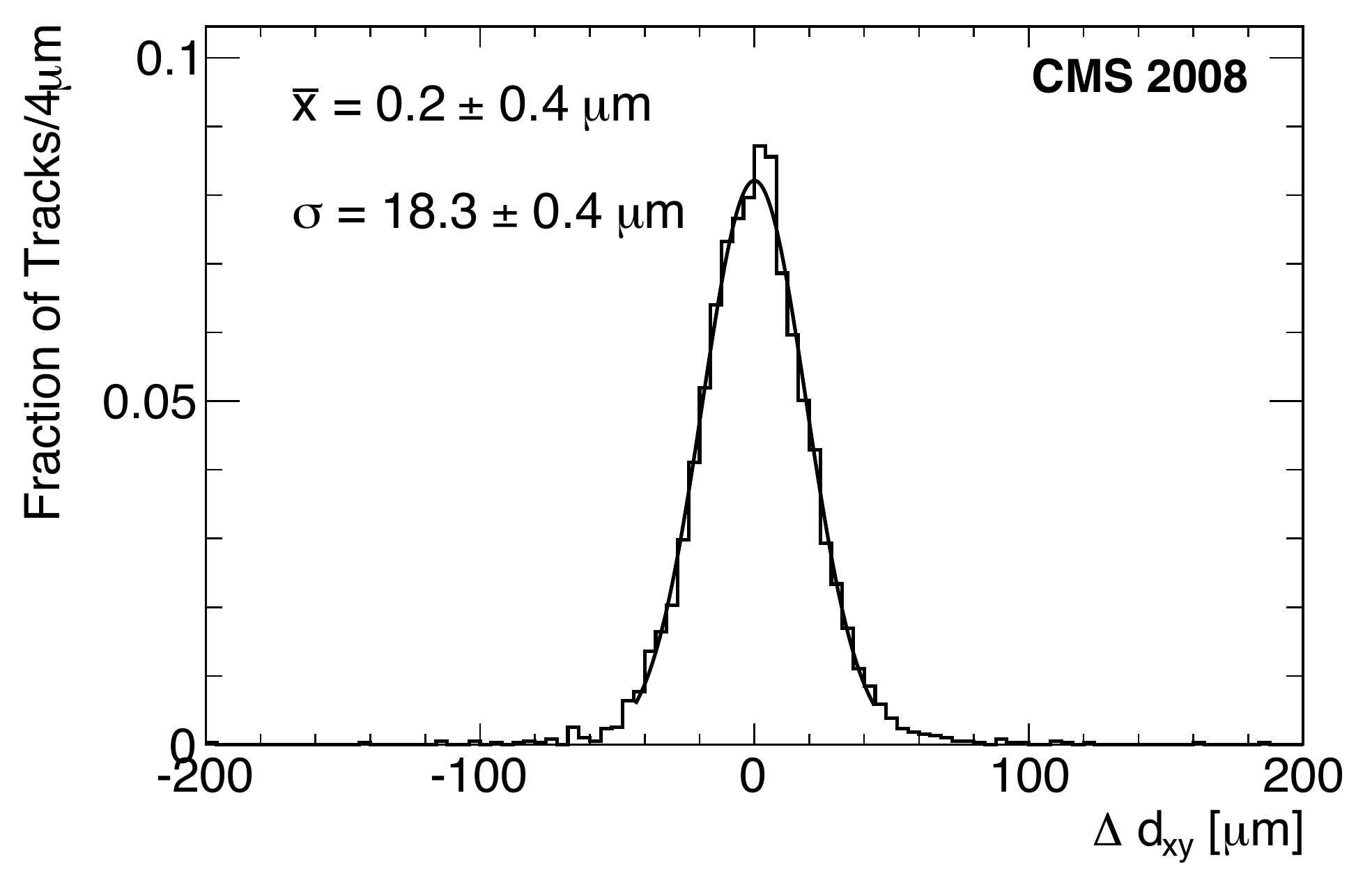}
	  \label{fig:dxy_residuals_20} 
      }}
    }
    \hspace{-0.5cm}
    \mbox{
      \subfigure[]
{\scalebox{0.47}{
	  \includegraphics[width=\linewidth]{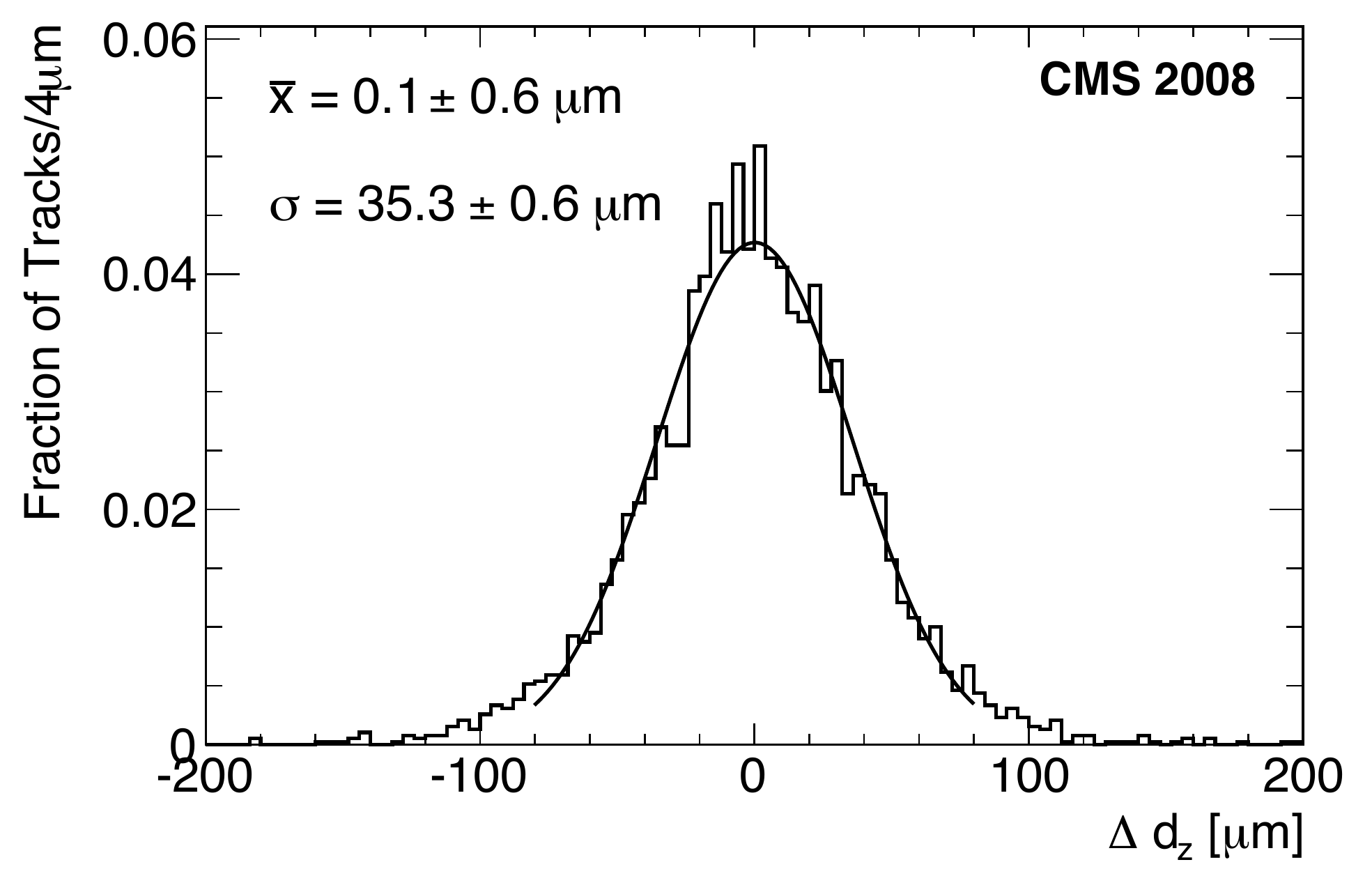}
	  \label{fig:dz_residuals_20} 
      }}
    }
    \caption{Residuals of the transverse (a) and longitudinal (b) track impact parameters measured with the track splitting method for track momentum larger than 4~GeV/$c$. The residuals for tracks with momentum larger than 20~GeV/$c$ are shown in (c) and (d) for the transverse and longitudinal plane, respectively. The solid line represents a Gaussian fit to the data.}
  \end{center}
\end{figure}

The distributions of the transverse ($d_{xy}$) and longitudinal ($d_z$) track impact parameter residuals are shown in Figs.~\ref{fig:dxy_residuals} and \ref{fig:dz_residuals}, respectively. The histogram shows the measurement and the solid line is a Gaussian fit to the data. The width of the residual distribution from the Gaussian fit is 23.3$\pm$0.3~$\mu$m and 39.4$\pm$0.5~$\mu$m in the transverse and longitudinal planes, respectively. 
If the track momentum is required to be greater than 20~GeV/$c$, the width of the residual distribution changes to 18.3$\pm$0.4~$\mu$m and 35.3$\pm$0.6~$\mu$m in the transverse and longitudinal planes, respectively, as shown in Figs.~\ref{fig:dxy_residuals_20} and \ref{fig:dz_residuals_20}. The momentum dependence of the impact parameter resolution is discussed in more detail in Ref.~\cite{alignment}.

\section{Summary\label{sec:conclusions}}

After installation in July 2008, the CMS silicon pixel detector was commissioned and then operated in a cosmic ray run in October--November 2008.  More than 96\% of the endcap disks and 99\% of the central barrel were included. About 270 million triggers were recorded, yielding about 85\,000 tracks traversing the pixel detector, which have been used to evaluate the performance of the detector.
While cosmic rays do not mimic beam collisions in terms of spatial distribution
and are asynchronous with respect to the LHC beam crossing clock,
these data allowed crucial testing of detector performance and of
calibration techniques.

Online charge calibration permitted a comparison of the charge measurements with the expectations from simulation, showing good agreement for the distribution widths, both in the barrel and endcap regions.  Initial operating thresholds of about 3000--3700 electrons per pixel were determined in calibration runs with charge injections.  Randomly arriving cosmic rays and the narrow time acceptance of the pixel ROC increase the effective threshold to approximately 5000 electrons, while a minimum ionizing particle deposits about 22\,000 electrons in the pixel sensors at normal incidence.

The Lorentz effect is crucial for the hit resolution in the CMS
pixel detector design, where Lorentz forces result in charge sharing
among pixels. The Lorentz shift was extracted from data and used to correct the reconstructed hit position. 
The measured values were found to be well reproduced by simulation studies, for both barrel and endcap regions.

Although limited by the narrow time acceptance of the front-end
electronics and the random arrival time of cosmic rays, a hit
efficiency study shows efficiencies greater than 96\% for most
detector modules.

Hit resolution measurements, even at this early stage of tuning, confirm the precision of the CMS pixel detector.  Using tracks that intersect overlapping barrel modules, hit resolutions of 19 $\mu$m ($r\phi$) and 31 $\mu$m ($z$) were extracted from the small event sample available.  The transverse (longitudinal) impact parameter resolution was found to be 18 $\mu$m (35 $\mu$m) for high momentum tracks, using a split-track technique. These results are in line with the expectations presented in Ref.~\cite{Bayatian:2006zz}. The CMS pixel detector will perform according to its specifications during operation with colliding beams.

\section*{Acknowledgements}

We thank the technical and administrative staff at CERN and other CMS Institutes, and acknowledge support from: FMSR (Austria); FNRS and FWO (Belgium); CNPq, CAPES, FAPERJ, 
and FAPESP (Brazil); MES (Bulgaria); CERN; CAS, MoST, and NSFC (China); COLCIENCIAS (Colombia); MSES (Croatia); RPF (Cyprus); Academy of Sciences and NICPB (Estonia); Academy 
of Finland, ME, and HIP (Finland); CEA and CNRS/IN2P3 (France); BMBF, DFG, and HGF (Germany); GSRT (Greece); OTKA and NKTH (Hungary); DAE and DST (India); IPM (Iran); SFI 
(Ireland); INFN (Italy); NRF (Korea); LAS (Lithuania); CINVESTAV, CONACYT, SEP, and UASLP-FAI (Mexico); PAEC (Pakistan); SCSR (Poland); FCT (Portugal); JINR (Armenia, Belarus, 
Georgia, Ukraine, Uzbekistan); MST and MAE (Russia); MSTDS (Serbia); MICINN and CPAN (Spain); Swiss Funding Agencies (Switzerland); NSC (Taipei); TUBITAK and TAEK (Turkey); 
STFC (United Kingdom); DOE and NSF (USA). Individuals have received support from the Marie-Curie IEF program (European Union); the Leventis Foundation; the A. P. Sloan 
Foundation; and the Alexander von Humboldt Foundation.

\bibliography{auto_generated}
\clearpage
\appendix
\cleardoublepage\appendix\section{The CMS Collaboration \label{app:collab}}\begin{sloppypar}\hyphenpenalty=500\textbf{Yerevan Physics Institute,  Yerevan,  Armenia}\\*[0pt]
S.~Chatrchyan, V.~Khachatryan, A.M.~Sirunyan
\vskip\cmsinstskip
\textbf{Institut f\"{u}r Hochenergiephysik der OeAW,  Wien,  Austria}\\*[0pt]
W.~Adam, B.~Arnold, H.~Bergauer, T.~Bergauer, M.~Dragicevic, M.~Eichberger, J.~Er\"{o}
% ...
\vskip\cmsinstskip
% ...

\end{sloppypar}
\end{document}